\documentclass[%
 reprint,
 amsmath,amssymb,
 aps,
prb,
]{revtex4-2}

\usepackage{graphicx}
\usepackage{dcolumn}
\usepackage{bm}
\usepackage[version=4]{mhchem} 
\usepackage{textcomp}
\usepackage{hyperref}
\usepackage[mathlines]{lineno}
\usepackage{braket} %
\usepackage{xcolor}
\usepackage{soul} 
\usepackage{color}
\usepackage{ulem}
\makeatletter

\newcommand{\Rmnum}[1]{\expandafter\@slowromancap\romannumeral #1@}
\makeatother
\hypersetup{
	colorlinks=true,
	linkcolor=blue,
	filecolor=blue,      
	urlcolor=blue,
	citecolor=blue,
    breaklinks=true,
    linktoc=all,     
}

\definecolor{highlight}{RGB}{255,255,153} 

\begin{document}


\title{Phaseless auxiliary-field quantum Monte Carlo method with spin-orbit coupling}

\author{Zheng Liu$^{1,2}$}
\author{Shiwei Zhang$^{3}$}
\author{Fengjie Ma$^{1,2}$}\email{fengjie.ma@bnu.edu.cn}

\affiliation{$^{1}$The Center for Advanced Quantum Studies and School of Physics and Astronomy, Beijing Normal University, Beijing 100875, China}
\affiliation{$^{2}$Key Laboratory of Multiscale Spin Physics (Ministry of Education), Beijing Normal University, Beijing 100875, China}
\affiliation{$^{3}$Center for Computational Quantum Physics, Flatiron Institute, New York, NY 10010, USA}

\date{\today}

\begin{abstract}
Spin-orbit coupling (SOC) is incorporated into the phaseless plane-wave-based auxiliary-field quantum Monte Carlo (pw-AFQMC) method. This integration is implemented using optimized multiple-projector norm-conserving pseudopotentials, which are derived from the fully-relativistic (FR) atomic all-electron Dirac-like equation. The inclusion of SOC enables accurate phaseless pw-AFQMC calculations that capture both electronic correlation and SOC effects concurrently, greatly improving the method’s applicability for studying systems containing heavy atoms. We discuss the form of FR pseudopotentials and detail the corresponding formulations of phaseless pw-AFQMC with a two-component Hamiltonian in the spinor basis. The accuracy of our approach is demonstrated by computing the dissociation energy of molecule \ce{I2} and the cohesive energy of bulk \ce{Pb}, highlighting the large influence of SOC in both. Subsequently, we determine the transition pressure of the  \Rmnum{3}-\Rmnum{5} compound \ce{InP} from its zinc-blende to rock-salt phase by constructing and analyzing their respective equations of state.
\end{abstract}


\maketitle


\section{\label{sec:level1}Introduction}

Spin-orbit coupling (SOC), the relativistic interaction between an electron’s spin angular momentum and its orbital motion in the nuclear electric field, is a fundamental quantum mechanical effect contributing to the electronic, magnetic, and topological properties of condensed matter systems \cite{Jackson, Tsymbal, RevModPhys.89.040502, RevModPhys.83.1057, RevModPhys.82.3045, RevModPhys.90.015001, 3m4m-3v59, RevModPhys.90.015005, RevModPhys.76.323, RevModPhys.82.53, RevModPhys.97.031001, RevModPhys.95.011002}. Its significance is particularly prominent in compounds containing transition-metal (3$d$, 4$d$, or 5$d$) and rare-earth (4$f$ or 5$f$) elements, where the large atomic number (\(Z\)) of constituent elements amplifies relativistic effects (scaling with \(Z^4\)) rendering SOC non-negligible for both fundamental quantum physics research and the development of emerging quantum devices \cite{PhysRevB.90.041112, 3m4m-3v59, RevModPhys.90.015005, RevModPhys.76.323, RevModPhys.82.53, RevModPhys.97.031001, RevModPhys.95.011002}. In such systems, SOC modulates magnetic exchange interactions, spin reorientation transitions, and magnetocrystalline anisotropy---key determinants of performance in permanent magnets and spintronic devices \cite{3m4m-3v59, RevModPhys.90.015005, RevModPhys.76.323, Xu2014}. Stronger SOC further gives rise to exotic quantum phenomena including topological insulators, Weyl semimetals, and spin-orbit Mott insulators, where the intricate interplay of SOC, electron correlation, and band topology yields unique electronic states \cite{RevModPhys.89.040502, RevModPhys.83.1057, RevModPhys.82.3045, RevModPhys.90.015001, PhysRevB.90.041112, science.1146006}. SOC additionally induces crystal field splitting and spin-lattice coupling, thereby enabling transformative applications in quantum sensors and optical quantum technologies \cite{PhysRevLett.132.190001, RevModPhys.89.035002, APR2025}. In all these cases, SOC is not a secondary perturbation but a core driver of material functionality; its omission or insufficient theoretical treatment invariably leads to erroneous predictions of a material’s structural, electronic, and magnetic properties. As demand surges for high-performance spintronic devices, topological quantum computing components, and next-generation magnetic storage materials, a precise physical understanding and accurate theoretical description of SOC in these complex systems have become indispensable prerequisites for advancing scientific frontiers and driving technological innovation.

To date, theoretical calculations for solid-state materials are predominantly based on the traditional density functional theory (DFT) \cite{DFT_Hohen,DFT_KS}. Such evaluations, which rely on various mean-field approximations, often break down for systems with strong electronic correlations, however. Many-body methods that rigorously account for electronic correlations are therefore highly desired \cite{science.aat5975, RevModPhys.73.33, RevModPhys.68.13, AFQMC}. While the explicit incorporation of SOC into standard DFT has been very mature and the associated calculations are widely adopted \cite{FR,NCPP_SRSO,NCPP_SOC,USPP_SOC}, evaluations with SOC based on many-body approaches pose formidable challenges. A primary bottleneck is the prohibitive computational cost: fully-relativistic (FR) calculations that include SOC drastically expand the effective Hilbert space of the system, amplifying the inherent computational expense of many-body methods. Scalar-relativistic (SR) \cite{SR} calculations are commonly employed, as they incur nearly identical computational costs to non-relativistic (NR) calculations yet provide more accurate descriptions of heavy atoms by incorporating relativistic kinematic effects (the mass-velocity and Darwin terms) while excluding the SOC interaction. However, an ever-growing body of research has uncovered a wealth of novel quantum phenomena that emerge from the increasingly strong SOC interactions in the strongly correlated materials. As a result, SR-level calculations are no longer sufficient, and the explicit inclusion of SOC, beyond mean-field approximations has become essential. This underscores the critical need for many-body methods that rigorously and accurately treat both effects in a high-precision manner.

The phaseless auxiliary-field quantum Monte Carlo (AFQMC) method \cite{AFQMC,Phase_AFQMC} is a powerful \textit{ab initio} many-body electronic structure approach with a suite of advantageous features. It mitigates the fermionic sign problem (or phase instability) via the phaseless approximation based on the trial wave functions, restores low-power (typically to the third power of system size) computational scaling, and supports the use of flexible one-particle basis sets. For the description of realistic systems, the phaseless AFQMC can be formulated within two canonical one-particle basis sets: the plane-wave basis set that is ubiquitous for extended condensed matter systems \cite{Phase_AFQMC,AFQMC_mol,AFQMC_press}, and the localized Gaussian-type basis set which is standard for atomic and molecular systems in quantum chemistry \cite{GTO_AFQMC1,GTO_AFQMC2}. One very important advantage of the plane-wave implementation with pseudopotentials is that its convergence to the basis-set limit depends uniquely and directly on the kinetic energy cutoff, enabling an unbiased representation of the wave function. The success of AFQMC method in the study of interacting many-fermion lattice models, e.g., the Hubbard model \cite{Hubbard}, highlights its capabilities for treating strong correlation. While the treatment of SOC in AFQMC has been explored for Gaussian-type orbitals \cite{GTO_SOC} and model Hamiltonians \cite{Model_SOC}, its integration into plane-wave-based AFQMC (pw-AFQMC) remains underdeveloped.

This work focuses on the incorporation of SOC into phaseless pw-AFQMC using optimized norm-conserving Vanderbilt (ONCV) \cite{Hamann} FR pseudopotentials, replacing $l$-dependent with $j$-dependent pseudopotentials for consistency with DFT procedures. The introduction of SOC requires the spins to explicitly enter the plane-wave basis which doubles the basis size, expanding the Hamiltonian into a $2\times2$ block matrix that mixes spins and applies to two-component spinor many-body wave functions. Related modifications for both the wave function propagation in the space of Slater determinants and the observable measurement with spinor plane-wave basis are discussed. This framework advances the many-body electronic structure method, enabling accurate treatment of both electron correlations and SOC, and extends its applicability to heavy-element-containing materials and bridges fundamental quantum physics and realistic material design. 

The remainder of this paper is organized as follows. In Sec.~\ref{sec:2}, we review the Hamiltonian by elaborating on each term and discuss the non-local component of pseudopotentials employed to incorporate SOC. Then we outline the theoretical framework and detailed implementations in phaseless pw-AFQMC. Section~\ref{sec:3} presents the applications of the developed SOC-included phaseless pw-AFQMC approach to validate the formalism and highlight the effects of SOC. We carry out calculations for the dissociation energy of simple molecule \ce{I2} and the cohesive energy of solid \ce{Pb}, by using SR and FR pseudopotentials, respectively. The results are also compared with both DFT calculations and the experiments. We then further investigate the zinc-blende to rock-salt structural transition of a \Rmnum{3}-\Rmnum{5} compound \ce{InP}. Finally, we summarize and conclude our work in Sec.~\ref{sec:4}.

\section{METHODOLOGY} \label{sec:2}

In this section, we begin by defining each term of the NR Hamiltonian in a plane-wave basis to establish the framework for subsequent discussions. We then elaborate on the forms of non-local pseudopotentials to incorporate SOC. Detailed implementations of phaseless pw-AFQMC, including propagation and measurement with SOC, are discussed following these.

\subsection{Hamiltonian}

The electronic Hamiltonian, separated from that of ions under the Born-Oppenheimer approximation \cite{BO_app}, can be expressed as 
\begin{equation}
\label{eq:H0}
    H = K + V_{\text{e-i}} + V_{\text{e-e}} + V_{\text{II}} ,
\end{equation}
where $K$, $V_{\text{e-i}}$, $V_{\text{e-e}}$, and $V_{\text{II}}$ are, respectively, the kinetic energy and electron-ion, electron-electron, and classical Coulomb ion-ion interactions \cite{Ion_ion}. The plane-wave basis (in real space representation),
\begin{equation}
    \langle{\mathbf{r}}\ket{\mathbf{k}+\mathbf{G}} = \bra{\mathbf{r}}c_{\mathbf{k}+\mathbf{G}}^\dagger\ket{0}
    = \frac{1}{\sqrt{\Omega}}e^{i(\mathbf{k}+\mathbf{G})\cdot\mathbf{r}},
\end{equation}
is usually adopted for extended systems with periodic boundary conditions. Here, $\Omega$ denotes the volume of the simulation cell, $\mathbf{G}$ is the reciprocal lattice vector, the vector $\mathbf{k}$ specifies a particular choice of periodic boundary condition for the system, and $c_{\mathbf{k}+\mathbf{G}}^\dagger$ is the corresponding creation operator. We define $M$ as the number of plane-waves within the kinetic energy cutoff and $N$ as the number of electrons in the system.

Concretely, each term of the Hamiltonian can be further expressed in second quantization as follows. The kinetic energy, expressed as 
\begin{equation}
\label{eq:3}
    K=\frac{1}{2}\sum_{\mathbf{G},\lambda} \mathbf{G}^2 c_{\mathbf{G},\lambda}^{\dagger}c_{\mathbf{G},\lambda}^{},
\end{equation}
and the pseudopotential that models the electron-ion interaction, given by 
\begin{equation}
    V_{\text{e-i}} = V^{\text{loc}} + V^{\text{nl}},
    \label{eq:4}
\end{equation}
where its local and non-local components take the forms 
\begin{subequations}
\label{eq:5}
\begin{equation}
    V^{\text{loc}} = \sum_{\lambda,\mathbf{G},\mathbf{G'}} V^{\text{loc}}(\mathbf{G}-\mathbf{G'})
    c_{\mathbf{G},\lambda}^{\dagger}c_{\mathbf{G'},\lambda}^{},
\end{equation}

\begin{equation}
    V^{\text{nl}} = \sum_{\lambda,\mathbf{G},\mathbf{G'}} V^{\text{nl}}(\mathbf{G},\mathbf{G'})
    c_{\mathbf{G},\lambda}^{\dagger}c_{\mathbf{G'},\lambda}^{},
\end{equation}
\end{subequations}
collectively form the one-body part of the Hamiltonian. We discuss details of $V^{\text{loc}}(\mathbf{G}-\mathbf{G'})$ and $V^{\text{nl}}(\mathbf{G},\mathbf{G'})$, which represent the matrix elements corresponding to the local and non-local components of the pseudopotential, respectively, in the subsequent subsection. Both reciprocal lattice vectors $\mathbf{G}$ and $\mathbf{G'}$ are restricted to the basis set $\{\mathbf{G}\}$ whose size is controlled by the kinetic energy cutoff $E_\text{cut}$. Additionally, the summation over the spin index $\lambda$ is written explicitly here. For brevity, any shift of the vector $\mathbf{k}$ is omitted, as all calculations are performed at a single fixed $\mathbf{k}$ point.

The electron-electron interaction term is
\begin{eqnarray}
\label{eq:6}
    V_{\text{e-e}} &=& \frac{1}{2}N\xi + \frac{1}{2}\frac{4\pi}{\Omega} \sum_{\mathbf{Q}\neq0}
    \frac{1}{Q^2}\rho^\dagger(\mathbf{Q})\rho(\mathbf{Q}) \nonumber\\
    && - \frac{1}{2}\frac{4\pi}{\Omega} \sum_{\lambda}\sum_{\mathbf{G} \neq \mathbf{G'}}
      \frac{1}{\vert \mathbf{G}-\mathbf{G'} \vert ^2} c_{\mathbf{G},\lambda}^{\dagger}c_{\mathbf{G},\lambda}^{},
\end{eqnarray}
where the constant $\xi$ stands for the self-interaction of an electron with its own periodic images evaluated via the standard Ewald summation method \cite{Ewald,Ewald_ori,Martin}. The third term, analogous to the kinetic energy term, corresponds to a sum of diagonal one-body operators arising from the anti-commutation of the fermion creation and annihilation operators. Here, we further introduce an important quantity, the one-body density operator
\begin{equation}
    \rho(\mathbf{Q}) = \sum_{\lambda,\mathbf{G}} 
    c_{\mathbf{G}-\mathbf{Q},\lambda}^{\dagger}c_{\mathbf{G},\lambda}^{}
    \theta(E_\text{cut}-\frac{1}{2} \vert \mathbf{G}-\mathbf{Q} \vert ^2),
\end{equation}
where $\mathbf{Q}=\mathbf{G}-\mathbf{G'}$, and the step function $\theta$ enforces that all $(\mathbf{G}-\mathbf{Q})'s$ fall within the plane-wave basis set. Correspondingly, the local potential can also be recast in terms of the one-body density operator.

It is worth emphasizing that all the aforementioned expressions are derived within the framework of a spin-unentangled Hamiltonian. In practical calculations, this allows us to work directly with the spatial plane-wave basis for each spin channel, thereby obviating the need for a rigorous spin-related expansion of the basis set.

\subsection{Pseudopotentials}

The adoption of pseudopotentials effectively eliminates core electronic states---those tightly bound to atomic nuclei, non-participating in chemical bond formation, and approximately invariant across atomic, molecular, and solid-state environments. This approximation simplifies the quantum mechanical problem to the exclusive treatment of valence electrons subject to an effective potential. Relativistic effects can be naturally incorporated into pseudopotentials, as their contribution to valence electrons arises predominantly in the nuclear-proximal interior of the atom. Widely used SR pseudopotentials are thus generated to include kinematic relativistic effects and effectively account for SOC by the construction of $j$-averaged pseudopotentials for each angular momentum $l$. While such an approximation yields band structures with no spin-orbit splitting and is broadly acceptable for many material electronic structure calculations, it becomes inadequate in cases where observable quantities---such as hole effective masses and spin relaxation times---stem directly from spin-orbit splitting. Kleinman has shown that for valence electron wave functions outside the core region, all relativistic effects up to second order of the fine-structure constant can be captured by solving a Pauli-like Schr\"odinger equation with pseudopotentials derived from FR radial all-electron atomic equations \cite{FR,Kleinman_FR}. We henceforth refer to them as FR pseudopotentials, in analogy to the NR and SR counterparts. We initiate our discussion by outlining the formalisms of NR and SR pseudopotentials.

The plane-wave-based matrix elements corresponding to the local potential component take the form
\begin{equation}
    V^{\text{loc}}(\mathbf{Q}) = \frac{1}{\Omega}\sum_{\gamma,\alpha}
    e^{-i\mathbf{Q}\cdot\mathbf{u}_{\gamma,\alpha}}V_\alpha^\text{loc}(\mathbf{Q}),
\end{equation}
where $\mathbf{u}_{\gamma,\alpha}$ is the position vector of the $\gamma$-th atom of species $\alpha$ within the simulation cell, measured relative to the cell's origin. The quantity $V_\alpha^\text{loc}(\mathbf{Q})$ represents the (unnormalized) Fourier transform of the atomic local potential.

The non-local component of the pseudopotential for a given atomic species can be expressed in the separable Kleinman-Bylander form \cite{KB} (non-local in both angular and radial variables) as
\begin{equation}
\label{eq:19}
    V_\alpha^\text{nl} = \sum_{l,m} \frac{ \ket{V_{\alpha l}\varphi_{\alpha l}^\text{ps}Y_{lm}} 
    \bra{Y_{lm}\varphi_{\alpha l}^\text{ps}V_{\alpha l}} }
    {\eta_{\alpha l}},
\end{equation}
where the pseudopotential $V_{\alpha l}$ and pseudo-orbital $\varphi_{\alpha l}$ rely only on the radial distance $r$ when represented in real space, and $Y_{lm}$ denotes the standard spherical harmonic function. The normalization factor, $\eta_{\alpha l}=\langle{Y_{lm}\varphi_{\alpha l}^\text{ps}} \ket{V_{\alpha l}|\varphi_{\alpha l}^\text{ps}Y_{lm}}$, is the overlap between pseudo-wavefunction and its corresponding projector $V_{\alpha l}\ket{\varphi_{\alpha l}^\text{ps}Y_{lm}}$. The overall matrix elements of the non-local pseudopotential take a fully separable form:
\begin{equation}
\label{eq:20}
    V^{\text{nl}}(\mathbf{G},\mathbf{G'}) = \sum_{j\in\{ \gamma,\alpha,l,m \} } \frac{1}{\eta_j}
    F_j^*(\mathbf{G})F_j(\mathbf{G'}),
\end{equation}
where we introduce
\begin{equation}
    F_j(\mathbf{G})=\frac{4\pi}{\sqrt{\Omega}}e^{i\mathbf{G}\cdot\mathbf{u}_{\gamma,\alpha}}
    f_{\alpha l}(G) Y_{lm}^*(\mathbf{G}).
\end{equation}
Here, $f_{\alpha l}(G)$ is the (radial component) Fourier transform of the projector, which can be read directly into our phaseless pw-AFQMC code.

Hamann generalized Eq.~(\ref{eq:19}) to the case of multiple projectors \cite{Hamann} based on the framework of Vanderbilt's ultrasoft pseudopotentials (USPPs) \cite{Van_USPP}. We accordingly rewrite the non-local pseudopotential component within the multi-projector formalism as 
\begin{equation}
\label{eq:23}
    V^\text{nl} = \sum_{\tau,l,m} B_{l,m}^\tau \ket{\chi_{\tau,l}Y_{l,m}}\bra{Y_{l,m}\chi_{\tau,l}},
\end{equation}
where $\ket{\chi_{\tau,l}}$ refers to the normalized form of the $\tau$-th projector function $V_{\tau,\alpha l}\ket{\varphi_{\tau,\alpha l}^\text{ps}}$ (with only radial part). The coefficients $B_{l,m}^\tau$ correspond to diagonal elements of $N_c$-dimensional (inverse) overlap matrix $B_{i,j}$, which is constructed from projector-pseudo-wavefunction inner products and subsequently diagonalized; here, the composite indices are defined as $i=\{\tau,l\}$ and $j=\{\tau',l'\}$, and $N_c$ is the number of total projectors, e.g., $N_c=8$ when $l_\text{max}=3$ and two projectors are assigned per $l$ during the pseudopotential generation. Note that these coefficients typically take the form of the inverse of $1/\tilde{b}_i$ (following the notation in Ref.~\cite{Hamann}) and differ from the  matrix $D_{i,j}$ introduced in Refs.~\cite{Van_USPP,USPP_SOC}, which is associated with the $\beta$ functions. We retain the magnetic quantum number $m$ in the subscript, albeit the coefficients are in fact independent of $m$: for a fixed $l$, all values of $m$ $(-l\leqslant m\leqslant l)$ share an identical $B_{l,m}^\tau$ for a given projector. Henceforth, we restrict our discussion to a single atom of a specified species; this simplification is justified by the fact that pseudopotential files directly provide these data (excluding the spherical harmonic component). The total non-local potential for the full system is then constructed as a simple sum of all atomic contributions, optionally weighted by the relevant atomic structure factors, as outlined in the preceding formalism.

An NR or SR pseudopotential constructed from the radial components of solutions to the corresponding-level equations depends exclusively on the orbital angular momentum $l$, as discussed above. The SOC effects omitted in the SR formalism can be incorporated by reformulating the pseudopotentials, with all other physical quantities remaining formally unchanged in expressions. In contrast, an FR pseudopotential depends on both the total angular momentum $j$ and the orbital angular momentum $l$. Specifically, upon coupling the orbital angular momentum to a spin angular momentum of $s=1/2$, distinct solutions arise for $j=l\pm1/2$ when $l>0$, whereas only $j=l+1/2$ exists for $l=0$. Analogous to Eq.~(\ref{eq:23}), the non-local operator (for each component) now takes the form of a ``$2\times2$'' matrix, given by
\begin{equation}
\label{eq:24}
    V^{\text{nl},\lambda,\lambda'} = \sum_{\tau,l,j,m_j} B_{l,j,m_j}^\tau
    \ket{\chi_{\tau,l,j}\widetilde{Y}_{l,j,m_j}^\lambda} \bra{\widetilde{Y}_{l,j,m_j}^{\lambda'}\chi_{\tau,l,j}},
\end{equation}
where $\widetilde{Y}_{l,j,m_j}$ are the eigenfunctions of the operator set $\{ L^2,S^2,J^2,J_z\}$ (compared to $Y_{l,m}$ in $\{ L^2,L_z\}$) and are commonly referred to as spin-angle functions. $\widetilde{Y}_{l,j,m_j}$ are spinor-valued, for instance, $j=l+1/2$ adopt the form
\begin{equation}
    \widetilde{Y}_{l,j,m_j} =
    \begin{pmatrix}
        \left(\dfrac{l+m+1}{2l+1}\right)^\frac{1}{2}Y_{l,m} \\
        \left(\dfrac{l-m}{2l+1}\right)^\frac{1}{2}Y_{l,m+1}
    \end{pmatrix},
\end{equation}
wherein $m=m_j-1/2$. It is the conventional expansion with respect to the spherical harmonics and spin wave function, and the pre-factors are so-called Clebsch-Gordan coefficients. The local potential remains the $j$-averaged form, such that the non-local operator defined in Eq.~(\ref{eq:24}) represents the sole component of the Hamiltonian responsible for spin mixing.

To incorporate SOC, we modify the pseudopotentials-reading module in our code to accommodate split $j$-dependent projectors and directly substitute spherical harmonics with spin-angle function to construct the non-local operator. The manner above, with a single projector, is basically equivalent to the formalism presented in Ref.~\cite{NCPP_SOC}. We note that alternative methodologies exist for treating the non-local operator in the presence of SOC. Our discussion of FR pseudopotentials above partially follows the organizational structure of Ref.~\cite{USPP_SOC}, wherein the coefficients of the non-local operator are first expressed in the general form
$(D_{\tau,l,j,m_j;\tau',l',j',m_{j'}} = D_{\tau,\tau'}^{l,j}\delta_{l,l'}\delta_{j,j'}\delta_{m_j,m_{j'}})$. Subsequently, a unitary rotation is performed on the spherical harmonics to recast Eq.~(\ref{eq:24}) into the form of Eq.~(\ref{eq:23}), at the expense of breaking the diagonality. Notably, this cost does not introduce any difficulty in the implementation of USPPs. This approach to handling FR pseudopotentials of various types is implemented in the \textit{Quantum ESPRESSO} code \cite{QE1,QE2}. By contrast, our implementation leverages the full diagonalities (including that between projectors) with only moderate modifications for existing multi-projector ONCV pseudopotentials in our phaseless pw-AFQMC framework \cite{AFQMC_ONCV}. One can also naturally decompose the full non-local potential into $j$-averaged SR and spin-orbit projectors while substituting spin-angle functions in terms of spherical harmonics \cite{NCPP_SRSO,Kleinman_FR,FR}, based on the relation $\mathbf{L\cdot\mathbf{S}}=l/2, -(l+1)/2$ for $j=l+1/2, l-1/2$, respectively. The ONCVPSP code \cite{ONCVPSP} supports the generations of both direct $j$-dependent projectors $\ket{\chi_{\tau,l,j}}$ as in Eq.~(\ref{eq:24}) (for \textit{Quantum ESPRESSO} as well as our phaseless pw-AFQMC) and of separate SR and spin-orbit components (for \textit{ABINIT} code \cite{Abinit}).

\subsection{AFQMC with SOC}

With the Hamiltonian formally established, we now proceed to detail the implementation of incorporating SOC into phaseless pw-AFQMC. To this end, we present and contrast theoretical frameworks and corresponding expressions with and without SOC. The formalism outlined below is universally applicable to other systems (or in a different basis set) within the AFQMC framework.

The ground-state wave function $\ket{\phi_0}$ is obtained by repeatedly applying the imaginary-time projection operator \cite{Proj} to an arbitrary trial wave function $\ket{\psi_\text{T}}$ that is non-orthogonal to $\ket{\phi_0}$, 
\begin{equation}
    \ket{\phi_0} = \lim_{n\rightarrow\infty}(e^{-\Delta\tau(H-E_\text{T})})^n \ket{\psi_\text{T}},
\end{equation}
where $E_\text{T}$ is an initial estimate of the ground-state energy typically derived  from $\ket{\psi_\text{T}}$. A small time-step $\Delta\tau$ facilitates the separation of one-body ($H_1$) and two-body ($H_2$) Hamiltonian components via second-order Trotter–Suzuki decomposition \cite{Trotter,Suzuki}: 
\begin{equation}
    e^{-\Delta\tau H} = e^{-\frac{1}{2}\Delta\tau H_1}e^{-\Delta\tau H_2}e^{-\frac{1}{2}\Delta\tau H_1} + 
    \mathcal{O}(\Delta\tau^3),
\end{equation}
wherein constant Hamiltonian terms have been omitted and excluded throughout the projection process. The error introduced by a finite $\Delta\tau$ can be systematically removed by extrapolation to the limit $\Delta\tau=0$. This decomposition scheme permits the independent treatment of each Hamiltonian component, streamlining both propagation and measurement, as well as the seamless incorporation of SOC.

\subsubsection{Propagation}

The propagation, namely the action of the one-body exponential propagator $B=e^{-\Delta\tau H_1}$ on a Slater determinant $\ket{\phi}$ (referred to as a walker), is operated via the multiplication of matrices $\Phi'=\rm{B}\Phi$. Note that the upright, uppercase symbols $\rm{B}$ and $\Phi$ denote the corresponding matrix representations of the propagator and walker, respectively. The resulting product matrix $\Phi'$ simply leads to another Slater determinant $\ket{\phi'}$, according to Thouless theorem \cite{Thouless1,Thouless2,Thouless_Ham}.

In the absence of SOC, the complete basis set can be expressible in terms of plane-waves with no spin dependence. The propagator matrix $\rm{B}$ is consequently in a dimension of $M \times M$. A random walker can be further factorized into two independent spin components, given by
\begin{equation}
\label{eq:10}
    \ket{\phi}=\ket{\phi^\uparrow}\otimes\ket{\phi^\downarrow},
\end{equation}
where $\Phi^\lambda$, the matrix form of $\ket{\phi^\lambda}$, is of size $M \times N_\lambda$: 
\begin{equation}
\label{eq:11}
    \begin{pmatrix}
        \varphi_{1,1}^\lambda & \varphi_{1,2}^\lambda & \cdots & \varphi_{1,N_\lambda}^\lambda \\
        \varphi_{2,1}^\lambda & \varphi_{2,2}^\lambda & \cdots & \varphi_{2,N_\lambda}^\lambda \\
        \vdots                & \vdots                & \ddots & \vdots                        \\
        \varphi_{M,1}^\lambda & \varphi_{M,2}^\lambda & \cdots & \varphi_{M,N_\lambda}^\lambda
    \end{pmatrix},
\end{equation}
with each column vector representing a single-particle orbital $\ket{\varphi}$. Here $\varphi_i$ are the coefficients in the expansion $\ket{\varphi}=\sum\nolimits_i^M\varphi_i\ket{\varepsilon_i}$ in terms of basis $\ket{\varepsilon_i}$, which is the plane-wave $\ket{\mathbf{k}+\mathbf{G}}$ here. We impose that the number of electrons for each spin component is $N_\lambda$ ($N_\uparrow+N_\downarrow=N$), since $S_z=N_\uparrow-N_\downarrow$ constitutes a good quantum number and can be fixed throughout the calculation.

A direct application of the two-body propagator is computationally infeasible. The Hubbard-Stratonovich (HS) transformation \cite{HS1,HS2} can therefore be employed to decompose the two-body propagator into a high-dimensional integral of one-body propagators over auxiliary fields
\begin{equation}
    e^{-\Delta\tau H_2}=(\frac{1}{\sqrt{2\pi}})^{D_{\boldsymbol{\sigma}}}
        \int d{\boldsymbol{\sigma}} e^{ -\frac{1}{2}{\boldsymbol{\sigma}}\cdot{\boldsymbol{\sigma}} }
        e^{ \sqrt{\Delta\tau}{\boldsymbol{\sigma}}\cdot{\mathbf{v}} },
\end{equation}
where the vector $\boldsymbol{\sigma}\equiv\{\sigma_i\}$ assembles all fields with the dimension $D_{\boldsymbol{\sigma}}$ given by the number of $\mathbf{Q}$ vectors allowable within the basis set. The one-body operator ${\mathbf{v}}$ is in linear combinations of the one-body density operator and its Hermite conjugate \cite{Phase_AFQMC,AFQMC_mol}.

Upon the incorporation of SOC, the primary modification is that the original $M$-dimensional plane-wave basis should be replaced by the generic basis, i.e., the two-component spinor plane-wave basis $\ket{\mathbf{k}+\mathbf{G}}\ket{\lambda}$ of dimension $2M$, in which $\ket{\lambda}$ denotes the spin eigenstate basis. A key consequence of SOC is that the spin components can no longer be decoupled, as the spin operator $S_z$ no longer commutes with the Hamiltonian. Each walker $\Phi$ representing a Slater determinant is now encoded as a $2M\times N$ matrix
\begin{equation}
    \begin{pmatrix}
        \varphi_{1,1} & \varphi_{1,2} & \cdots & \varphi_{1,N} \\
        \varphi_{2,1} & \varphi_{2,2} & \cdots & \varphi_{2,N} \\
        \vdots        & \vdots        & \ddots & \vdots        \\
        \varphi_{2M,1} & \varphi_{2M,2} & \cdots & \varphi_{2M,N}
    \end{pmatrix}.
\end{equation}
By partitioning this matrix according to the ``up'' and ``down'' components in the spinor basis, the representation can be compactly rewritten in block form as 
\begin{equation}
\label{eq:27}
    \Phi=
    \begin{pmatrix}
        \Phi^\uparrow \\
        \Phi^\downarrow
    \end{pmatrix}.
\end{equation}
This block structure directly corresponds to our practical implementation strategy, wherein the matrix is stored by appending an additional spin dimension to the basis indexing. It is important to note that the submatrix $\Phi^\lambda$ (each of size $M\times N$) now refers to only a component in the full Slater determinant. This stands in contrast to their interpretation in Eqs.~(\ref{eq:10}) and~(\ref{eq:11}) (in the non-SOC case), where $\Phi^\lambda$ represented the $M\times N_\lambda$ wave function matrix associated exclusively with spin-$\lambda$ electrons.

Accordingly, each one-body propagator $\text{B}$ (including those decompositions from the two-body Hamiltonian component) is of size $2M\times2M$ and will be generalized as
\begin{equation}
\label{eq:28}
    \mathrm{B} = 
    \begin{pmatrix}
        \mathrm{B}^{\uparrow\uparrow} & \mathrm{B}^{\uparrow\downarrow} \\
        \mathrm{B}^{\downarrow\uparrow} & \mathrm{B}^{\downarrow\downarrow}
    \end{pmatrix}.
\end{equation}
For propagators that exhibit no spin-mixing in the pure plane-wave basis, their original $M\times M$ matrix representation $\mathrm{B}_M$ populates the diagonal blocks 
$\mathrm{B}^{\uparrow\uparrow}$ and $\mathrm{B}^{\downarrow\downarrow}$. As established earlier, the non-local component of the pseudopotential is the only Hamiltonian term that couples spins, implying that the off-diagonal blocks satisfy 
$\mathrm{B}^{\uparrow\downarrow}=\mathrm{B}^{\downarrow\uparrow}=0$ for all other one-body terms. Given that each Hamiltonian component is treated independently during the propagation and measurement, terms lacking spin-mixing can retain their original non-relativistic forms without modifications. For instance, a single propagation step for these spin-diagonal terms can be realized by 
\begin{equation}
\label{eq:29}
    \Phi^{'\lambda} = \mathrm{B}_M \Phi^{\lambda},
\end{equation}
thereby circumventing the need to construct and manipulate a full $2M\times2M$ matrix.

In practice, we do not even store full $M\times M$ matrices but instead exploit more efficient schemes, as direct matrix storage and multiplication incur prohibitive costs, particularly for the large basis set size of plane-waves. The kinetic energy in Eq.~(\ref{eq:3}) and the exchange term in Eq.~(\ref{eq:6}) are both diagonal, such that their application to the wave function reduces to simple one-dimensional array multiplications executed within a particle loop. Furthermore, fast Fourier transform (FFT) is regularly invoked for both evaluation of matrix elements and matrix multiplication, thanks to the nature of plane-waves. For example, residual one-body operators arising from the HS transformation of two-body interactions, as well as the local pseudopotential operator, can be expressed in terms of the one-body density matrix $\rho(\mathbf{Q})$ and propagated based on 
\begin{eqnarray}
\label{eq:13}
    &&\exp{ \left[ {\sum_\mathbf{Q}\sqrt{\frac{\Delta\tau}{Q^2}} f(\mathbf{Q})\rho(\mathbf{Q})} \right] }
    \ket{\phi} \nonumber \\
    && \simeq \sum_{n=0}^{n_\text{max}} \frac{1}{n!} \left[ \sum_\mathbf{Q}
    \sqrt{\frac{\Delta\tau}{Q^2}}f(\mathbf{Q})\rho(\mathbf{Q}) \right]^n \ket{\phi},
\end{eqnarray}
where each term in the series can be computed via iterative FFT operations, resulting in accelerated propagations. A truncation threshold of $n_\text{max}\simeq4$ is sufficient to reproduce the propagator with high fidelity for typical values of $\Delta\tau$ \cite{AFQMC_mol}.

The non-local operator, however, presents greater challenges due to its intrinsic non-locality. We begin by taking the original $j$-averaged non-local operator as an illustration. As specified in Eq.~(\ref{eq:20}), its matrix elements are obtained from products of $F_j(\mathbf{G})$, whose dimension is $N_j\times M$. Here, size $N_j$ is determined by the composite index $\{ \gamma,\alpha,l,m \}$. The $M\times M$ non-local potential matrix can thus be expressed as $\mathrm{V}_\text{nl}=\mathrm{E}\mathrm{F}$, where the matrix elements of $\mathrm{E}$ and $\mathrm{F}$ correspond to $F_j^*(\mathbf{G})/\eta_j$ and $F_j(\mathbf{G})$, respectively. Its propagator is expanded as
\begin{eqnarray}
\label{eq:30}
    \mathrm{exp}\left(-\frac{\Delta\tau}{2}\mathrm{V}_\text{nl}\right) = \sum_{n=0} \frac{1}{n!}
    \left( -\frac{\Delta\tau}{2}\mathrm{V}_\text{nl} \right)^n \nonumber \\
    = \mathrm{I} + \mathrm{E} \left[
    \sum_{n=1} \frac{1}{n!} \left(-\frac{\Delta\tau}{2}\right)^n \mathrm{A}^{n-1}
    \right] \mathrm{F},
\end{eqnarray}
where $\mathrm{I}$ is the identity matrix and $\mathrm{A}=\mathrm{FE}$ is a compact matrix of size $N_j\times N_j$. (Usually we sum up to $n=20$.) This replacement from $\mathrm{EF}$ $(M\times M)$ to $\mathrm{FE}$ $(N_j\times N_j)$ 
yields substantial savings in both storage and computational cost, as it obviates the need to directly exponentiate the large $M\times M$ matrix or apply it to the walker. In the propagation step, we compute the product $\mathrm{F}\Phi$ first to leverage this dimensional reduction.

Following the incorporation of SOC into the non-local operator (cf. Eq.~(\ref{eq:24})), its matrix elements take a form analogous to Eq.~(\ref{eq:20}) but now carry an explicit spin dependence. This spin dependence is encoded as 
\begin{equation}
    F_J(\mathbf{G,\lambda}) = \frac{4\pi}{\sqrt{\Omega}}e^{i\mathbf{G}\cdot\mathbf{u}_{\gamma,\alpha}}
    f_{\alpha,l,j}(G) \widetilde{Y}_{l,j,m_j}^{\lambda*},
\end{equation}
where the composite index $J$ is extended to $\{ \gamma,\alpha,l,j,m_j\}$. The construction of the SOC-inclusive propagator follows the same procedure as that outlined in Eq.~(\ref{eq:30}), with the key distinction that the matrix blocks (each of size $M\times M$) are populated according to four distinct spin configurations. The propagation is then executed within a spin-resolved loop.

\subsubsection{Measurement}

Observable measurement, performed on the converged ground-state wave function, in conjunction with the propagation, form the complete phaseless pw-AFQMC algorithm. The mixed estimator is always exploited to calculate the ground-state energy
\begin{equation}
    E_0=\frac{\bra{\psi_\text{T}}H\ket{\phi}} {\langle{\psi_\text{T}}\ket{\phi}}.
\end{equation}
Measurements are carried out via evaluating the Green's function \cite{AFQMC_Green} 
\begin{equation}
    G_{ji} = \frac{\bra{\psi}c_{i}^{\dagger}c_{j}\ket{\phi}} {\langle{\psi}\ket{\phi}}
    = \left[ \Phi(\Psi^\dagger\Phi)^{-1}\Psi^\dagger \right]_{ji}.
\end{equation}
In standard implementations, the state $\bra{\psi}$ is taken as the trial wave function $\bra{\psi_\text{T}}$; however, it can be replaced by any arbitrary Slater determinant, such as in back-propagation schemes \cite{AFQMC_Green,AFQMC_BP,PWAFQMC_BP_Den}.

For any one-body operator, for instance the kinetic energy, its estimator can be expressed as
\begin{equation}
    \braket{K} = \frac{1}{2}\sum_{\lambda}\sum_i^M \mathbf{G}_i^2 G_{ii}^{\lambda\lambda},
\end{equation}
where in general
$G_{ji}^{\lambda\lambda'}=\bra{\psi_\text{T}}c_{i,\lambda'}^{\dagger}c_{j,\lambda}\ket{\phi} / 
{\langle{\psi_\text{T}}\ket{\phi}}$
are the elements of this $2M\times2M$ matrix, which can be divided into four spin-resolved subblocks corresponding to all spin index combinations. For example, the upper-right $(1,2)$ subblock contains the matrix elements associated with $\uparrow\downarrow$ spin-flip processes. 

By applying Wick's theorem \cite{Santos_Wick}, the two-body electron-electron energy is estimated by
\begin{equation}
    \braket{V_\text{e-e}}=\frac{1}{2}\sum_{\lambda\lambda'}\sum_{ijkl}^{M}V_{\text{ee}}^{ijkl}
    (G_{il}^{\lambda\lambda}G_{jk}^{\lambda'\lambda'}-G_{ik}^{\lambda\lambda'}G_{jl}^{\lambda'\lambda}),
\end{equation}
where $V_\text{ee}^{ijkl}$ denotes the two-body matrix elements. The first term in parentheses corresponds to the direct (Coulomb) term, and the second to the cross (exchange) term.

In the absence of SOC, there are no spin-flip operators of the form 
$c_{i,\lambda'}^{\dagger}c_{j,\lambda}$ for $\lambda\neq\lambda'$, and Green’s function $G_{ij}^{\lambda\lambda'}=0$ for $\lambda\neq\lambda'$. In other words, the off-diagonal parts vanish rendering the matrix block-diagonal. The cross term is consequently reduced to a single spin index summation. Moreover, since the wave function is factorized into independent spin-up and spin-down components, the Green’s function can be computed by 
$G_{ji}^{\lambda} = \left[ \Phi^\lambda(\Psi^{\lambda\dagger}\Phi^\lambda)^{-1}\Psi^{\lambda\dagger} \right]_{ji} = G_{ji}^{\lambda\lambda}$.

It is natural to decompose the Green’s function and pre-compute 
\begin{equation}
    \Theta_{jq}^\lambda = \left[ \Phi^\lambda\left( \Psi^{\lambda\dagger}\Phi^\lambda \right)^{-1} \right]_{jq},
\end{equation}
which possesses the same dimensionality as a random walker matrix. The Green’s function then takes the form
\begin{equation}
    G_{ji}^\lambda=\sum_q^{N_\lambda} \Theta_{jq}^\lambda \Psi_{iq}^{\lambda*},
\end{equation}
where $i$ and $j$ index the basis vectors and $q$ denotes the particle index. For subsequent calculations, we additionally define the real-space representations via (inverse) Fourier transformation:
$\widetilde{\Theta}_{\mathbf{r}q}^\lambda =
\sum_je^{i\mathbf{G_j}\cdot{\mathbf{r}}}\Theta_{jq}^\lambda$ and
$\widetilde{\Psi}_{\mathbf{r}q}^{\lambda*} = \sum_je^{-i\mathbf{G_j}\cdot{\mathbf{r}}}\Psi_{jq}^\lambda$. (A tilde is affixed to the symbols here to denote their real-space form.)

We address measurement by detailing the mixed-estimator of the cross term in this part. The matrix elements of the two-body term in the plane-wave basis are given by 
\begin{eqnarray}
    V_{\text{ee}}^{ijkl} = \bra{\mathbf{k}+\mathbf{G}_i,\mathbf{k}+\mathbf{G}_j}V_{\text{e-e}} \ket{\mathbf{k}+\mathbf{G}_l,\mathbf{k}+\mathbf{G}_k} = 
    \frac{4\pi}{\Omega} \nonumber \\ \sum_\mathbf{Q}\frac{1}{Q^2}
    \int\frac{d\mathbf{r}}{\Omega} e^{-i(\mathbf{Q}+\mathbf{G}_l-\mathbf{G}_i)\cdot\mathbf{r}} 
    \int\frac{d\mathbf{r}'}{\Omega} e^{i(\mathbf{Q}-\mathbf{G}_k+\mathbf{G}_j)\cdot\mathbf{r}'}.
\end{eqnarray}
The wave vector $\mathbf{k}$ is explicitly retained in this expression; as with the local pseudopotential matrix elements, however, it does not play a role here. The mixed estimator of the cross term is then
\begin{eqnarray}
\label{eq:35}
    \langle V_{\text{e-e}}^{\text{ex}} \rangle &=& \frac{1}{2} \frac{4\pi}{\Omega} \sum_\mathbf{Q}\frac{1}{Q^2}
    \sum_\lambda\sum_{qq'}
    \int\frac{d\mathbf{r}}{\Omega} e^{-i\mathbf{Q}\cdot\mathbf{r}}
    \widetilde{\Theta}_{\mathbf{r}q}^\lambda \widetilde{\Psi}_{\mathbf{r}q'}^{\lambda*} \nonumber \\
    && \int\frac{d\mathbf{r'}}{\Omega} e^{i\mathbf{Q}\cdot\mathbf{r'}}
    \widetilde{\Theta}_{\mathbf{r'}q'}^\lambda \widetilde{\Psi}_{\mathbf{r'}q}^{\lambda*}.
\end{eqnarray}
By introducing $\widetilde{F}_x^\lambda(\mathbf{r},q,q') = \widetilde{\Theta}_{\mathbf{r}q}^\lambda \widetilde{\Psi}_{\mathbf{r}q'}^{\lambda*}$ and its Fourier transform $F_x^\lambda(\mathbf{Q},q,q')$, we can decompose the cross term into diagonal ($q=q'$) and off-diagonal ($q\neq q'$) contributions, followed by straightforward summations over the particle and spin indices.

The direct term is more straightforward to evaluate. As its formulation involves no index exchange or entanglement in the Green’s function, and the summation runs over the two spins independently, we can compute it solely by evaluating
$\widetilde{F}_d(\mathbf{r}) = \sum_{\lambda,q} \widetilde{\Theta}_{\mathbf{r}q}^\lambda
\widetilde{\Psi}_{\mathbf{r}q}^{\lambda*}$ (and its corresponding Fourier transform).

Accordingly, we can always treat spin-up and spin-down independently in both wave function propagation and observable measurement steps to perform spin-polarized or spin-unpolarized (including restricted case, i.e., orbitals with minority spin are identical to those with majority spin) calculations.

In the spinor basis, original spin-independent terms in the Hamiltonian remain block-diagonal and can be propagated separately for ``up'' and ``down'', as prescribed by Eq.~(\ref{eq:29}). The Green's function in identical form as Eq.~(\ref{eq:28}), however, is always a full $2M\times2M$ matrix due to the generic walker being represented as a $2M\times N$ matrix in the spinor basis. Our preceding implementation framework can be preserved with only minor modifications via the redefinition 
\begin{equation}
    \Theta_{jq}^\lambda = \left[ \Phi^\lambda\left( \Psi^{\dagger}\Phi \right)^{-1} \right]_{jq},
\end{equation}
where $\Phi^\lambda$ is defined as in Eq.~(\ref{eq:27}). The cross term to the two-body electron-electron energy and the non-local potential with SOC included require additional reformulations, while the estimation of all other terms in the Hamiltonian follows identical procedures as before. Since the cross term now involves a double summation over spin indices, we can somewhat simplify Eq.~(\ref{eq:35}) by introducing 
\begin{equation}
    \widetilde{F}_x(\mathbf{r},q,q') = \sum_\lambda \widetilde{\Theta}_{\mathbf{r}q}^\lambda \widetilde{\Psi}_{\mathbf{r}q'}^{\lambda*},
\end{equation}
in which the spin summation is performed internally. The non-local operators involve a double summation over spins (combined with basis sets) as well and can be estimated individually in a similar manner. 

All other relevant quantities in the simulation, e.g., the propagation of the weight, force bias, and mean-field background \cite{AFQMC_press,Phase_AFQMC}, can be evaluated using estimators constructed following the procedures outlined above.

\section{APPLICATIONS} \label{sec:3}

To elucidate the effects of SOC, a clean and instructive example is to calculate the relative binding energies of simple atomic or molecular systems, specifically for heavy $p$-block elements \cite{SOC_atom_mol,SOC_element,HalogenSOC}, with and without SOC included, respectively. In this section, we first apply our phaseless pw-AFQMC method to two benchmark cases: the dissociation energy of molecule \ce{I_2} and the cohesive energy of crystalline \ce{Pb}. Our results are benchmarked against available experimental data and compared with corresponding DFT calculations for reference. We then employ phaseless pw-AFQMC with SOC to compute the equations of state (EOS) for the zinc-blende (ZB) and rock-salt (RS) phases of the \Rmnum{3}-\Rmnum{5} compound \ce{InP}, and determine the phase transition pressure between these two phases. All DFT calculations were performed using the \textit{Quantum ESPRESSO} package \cite{QE1,QE2}.

\subsection{\ce{I_2} and \ce{Pb}}

Halogen elements with high atomic number are well known to exhibit strong SOC effects \cite{HalogenSOC}, e.g., post-$d$ elements \ce{Br} and \ce{I}, which have been widely studied in chemistry. We take the molecule \ce{I_2} as an illustration and calculate its dissociation energy as $D_e=2E_\text{atom}-E_\text{mol}$, where $E_\text{atom}$ and $E_\text{mol}$ are the ground state energies of the isolated atom and molecule, respectively. The molecular energy of \ce{I_2} is evaluated at the experimental equilibrium bond length of $2.666$ \AA~\cite{I2_De}. For all plane-wave-based calculations, we employ (SR and FR) local density approximation (LDA) pseudopotentials, generated from the atomic reference state [Pd]$5s^25p^5$ using the ONCVPSP code \cite{ONCVPSP}. Both DFT and phaseless pw-AFQMC calculations for the iodine atom and \ce{I_2} molecule are performed at the single $\mathbf{k}=\Gamma$ point, within a large cubic simulation cell of size $L=12$ \AA. In phaseless pw-AFQMC calculations, the trial wave function adopts the simplest form, of the corresponding DFT-LDA ground-state wave function.

\begin{figure}
    \centering
    \includegraphics[width=1.0\linewidth]{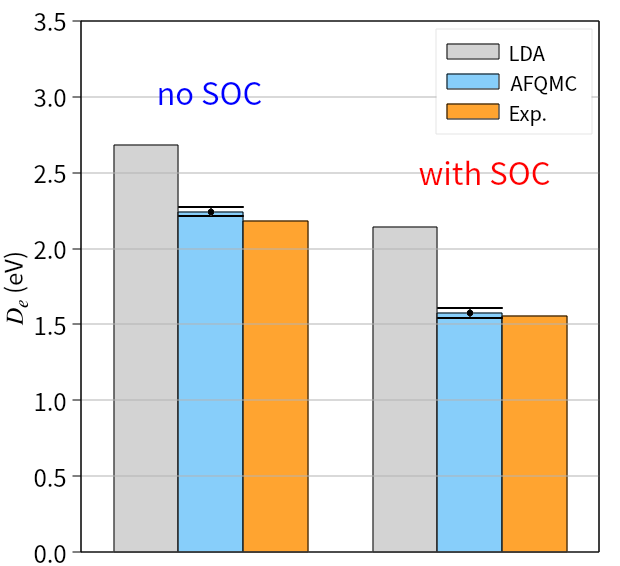}
    \caption{Dissociation energy of molecule \ce{I_2} from SR (labeled ``no SOC'') and FR (labeled ``with SOC'') calculations, comparing the results from DFT-LDA (gray bins), phaseless pw-AFQMC (blue bins) and experiment (orange bins). Statistical uncertainties for the phaseless pw-AFQMC calculations are represented by black horizontal error bars.}
    \label{De}
\end{figure}

The results with and without SOC from DFT-LDA and phaseless pw-AFQMC are presented in Fig.~\ref{De}, where the postprocessing finite-size correction scheme \cite{FS_corr,FS_mag_corr} is applied to both atomic and molecular energies from the phaseless pw-AFQMC calculations (although these corrections nearly cancel out in the $D_e$). The experimental reference bar for the non-SOC case (left-hand cluster) is obtained by approximate removal of SOC effects \cite{I2_SR} from the measured experimental dissociation energy \cite{I2_De}, and is included here for direct comparison. Compared to this ``$j$-averaged'' experimental value, non-SOC DFT-LDA and phaseless pw-AFQMC calculations overestimate the dissociation energy by $0.50$ and $0.06(3)$ eV, respectively. After the inclusion of SOC, DFT-LDA overestimates the energy by $0.59$ eV, while phaseless pw-AFQMC result is in excellent agreement with the experimental value within the statistical uncertainty. Our phaseless pw-AFQMC calculations of \ce{I_2} also yield results that are in accord with those reported in Ref.~\cite{GTO_SOC} at both relativistic levels, where the latter study was performed using a Gaussian basis set.

For atom \ce{I} in a spin-polarized calculation without SOC, it has an open-shell electronic configuration $5s^25p^5$ with $4\uparrow$ and $3\downarrow$ electrons, implying that a degeneracy exists for the $2\downarrow$ electrons in the $p$-orbitals. This degeneracy is treated via fractional occupations in DFT-LDA calculations, with $2/3$ of a $\downarrow$ electron assigned to each $p$-orbital. For phaseless pw-AFQMC calculations, in contrast, initializing from a single-determinant trial wave function---for instance, partially populating two of the three $p$-orbitals with the $2\downarrow$ electrons---explicitly breaks the three-fold symmetry of the $p$-orbital manifold. Additionally, the phaseless approximation, which uses the trial wave function to guide walkers' evolution, may compromise the numerical accuracy of the results to a certain extent. To explore this potential issue in greater depth, we constructed a more symmetric multi-determinant trial wave function that incorporates all three $p$-orbitals, generated by equally weighting linear combinations of any two $p$-orbitals for the $2\downarrow$ electrons. As discussed in Ref.~\cite{AFQMC_mol}, however, the resulting phaseless pw-AFQMC results show no appreciable deviations and agree with those from the single-determinant approach within statistical error bars. In other words, here the multi-determinant trial wave function presents results of equivalent quality to the single-determinant trial wave function.

We then turn our attention to bulk materials, of another substantially heavier element Pb with a large atomic number ($Z=82$). Here we focus on the cohesive energy defined as $E_\text{coh}=E_\text{atom}-E_\text{sol}$, which is obtained by calculating the ground-state energy of an isolated Pb atom ($E_\text{atom}$) and the per-atom ground-state energy of solid Pb ($E_\text{sol}$). For both SR and FR calculations, non-local LDA-type pseudopotentials are employed, with the $5d^{10}6s^26p^2$ electrons treated as valence states \cite{SG15}. The isolated Pb atom is simulated within a cubic supercell of $L=10$ \AA, while bulk face-centered cubic (fcc) Pb is modeled using the experimental lattice constant $a_0=4.95$ \AA.

\begin{figure}
    \centering
    \includegraphics[width=1.0\linewidth]{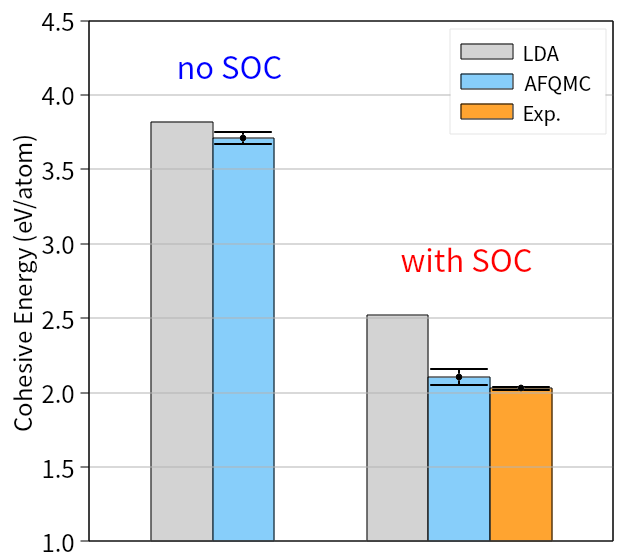}
    \caption{Cohesive energy of bulk Pb. DFT-LDA and phaseless pw-AFQMC results are displayed as gray and blue bins, respectively. Finite-size correction has been applied to the phaseless pw-AFQMC results. The orange bin with an error bar represents the experimental value with its associated uncertainty. Black horizontal bars on blue bins represent the stochastic uncertainties of the phaseless pw-AFQMC calculations.}
    \label{Coh}
\end{figure}

Fig.~\ref{Coh} presents the cohesive energy results from the DFT-LDA and phaseless pw-AFQMC calculations, with the experimentally measured value denoted in the right-hand cluster \cite{Pb_coh}. In phaseless pw-AFQMC calculations, we use the hybrid formalism \cite{Phase_AFQMC,AFQMC_press} for both the atom and bulk crystal, where a $4$-atom cubic supercell is adopted for bulk Pb. A single $\Gamma$ $\mathbf{k}$-point is used for the atom in both DFT-LDA and phaseless pw-AFQMC calculations. For fcc bulk Pb, the DFT-LDA calculations sample the Brillouin zone using a dense $16\times16\times16$ Monkhorst-Pack $\mathbf{k}$ grid \cite{MP} for the primitive cell, whereas phaseless pw-AFQMC calculations employ a single representative Baldereschi mean-value $\mathbf{k}$ point \cite{Bald} for cubic cell. Finite-size error corrections accounting for one-body and two-body contributions are additionally applied to the phaseless pw-AFQMC results. In the SR case without SOC, both DFT-LDA ($3.82$ eV) and phaseless pw-AFQMC ($3.71(4)$ eV) overestimate the experimental cohesive energy and the corresponding results including SOC by over $1$ eV, confirming the inherently strong SOC effects in Pb as anticipated. Upon the inclusion of SOC, the DFT-LDA value deviates from the experimental value by about $0.49$ eV, while the phaseless pw-AFQMC calculation reduces this discrepancy to a mere $0.07(5)$ eV.

While the good agreement between the phaseless pw-AFQMC results and experimental measurements is encouraging, our present treatment of bulk Pb adopts several simplifications. First, a larger supercell would be necessary to minimize finite-size errors as thoroughly as possible. Second, due to the metallicity of solid Pb, the use of twist-averaged boundary conditions instead of a single Baldereschi point would be more appropriate. These two factors are expected to largely counteract one another. Additionally, the experimental lattice constant and cohesive energy values used in this work are taken from different references, which may introduce a small systematic bias as well. Importantly, none of these simplifications should affect the subsequent analysis of SOC effects presented in this study. In DFT-LDA calculations, the inclusion of SOC reduces the cohesive energy by about $1.30$ eV, whereas this reduction amounts to as much as $1.61(7)$ eV in phaseless pw-AFQMC calculations. This disparity uncovers the intricate interplay between electronic correlation and SOC effects in Pb.

\subsection{Transition pressure of \ce{InP}}

\begin{table*}
\caption{\label{tab:table1}Structural properties of the ZB and RS phases of \ce{InP}, along with the transition pressure $P_t$ (GPa) between the two phases. Here, $a_0$ (\AA) is the equilibrium lattice constant and $B_0$ (GPa) is the zero-pressure bulk modulus. Theoretical works with LDA and GGA exchange-correlation functionals from other references are based on all-electron calculations. Also presented for comparison are our DFT results obtained using the same pseudopotentials as those employed in phaseless pw-AFQMC (SOC-AFQMC) calculations. Fitting uncertainties for the phaseless pw-AFQMC calculations are given in parentheses.}
\renewcommand{\arraystretch}{1.2}
\begin{ruledtabular}
\begin{tabular}{cccccc}
 &\multicolumn{2}{c}{ZB structure} &\multicolumn{2}{c}{RS structure} \\
& $a_0$ (\AA) & $B_0$ (GPa) & $a_0$ (\AA) & $B_0$ (GPa) & $P_t$ (GPa)\\ \hline
Experiment & $5.869$\footnotemark[1], $5.858$\footnotemark[6]
           & $76(4)$\footnotemark[2], $72.0$\footnotemark[6]      & &
           & $9.8(5)$\footnotemark[3], $10.8$\footnotemark[2] \\
LDA        & $5.838$\footnotemark[4]    & $72.06$\footnotemark[4]
           & $5.428$\footnotemark[4]    & $92.06$\footnotemark[4]  & $6.37$\footnotemark[4] \\
GGA        & $6.00$\footnotemark[5]    & $62.39$\footnotemark[5]
           & $5.55$\footnotemark[5]    & $79.12$\footnotemark[5]  & $7.35$\footnotemark[5],
           $8.90$\footnotemark[4]     \\
GGA (This work)        & $5.948$    & $58.0$    & $5.538$    & $75.2$  & $7.60$     \\
SOC-AFQMC  & $5.887(1)$ & $72.0(4)$ & $5.488(7)$ & $87(3)$ & $8.85(17)$ \\

\end{tabular}
\end{ruledtabular}
\footnotetext[1]{Reference \cite{Cry_Wyckoff}.}
\footnotetext[2]{Reference \cite{InP_Pt}.}
\footnotetext[3]{Reference \cite{4_35}.}
\footnotetext[4]{Reference \cite{InP_Pt_DFT1}.}
\footnotetext[5]{Reference \cite{InP_Pt_DFT2}.}
\footnotetext[6]{Reference \cite{HSE}. Zero-point effects are removed.}

\end{table*}

\begin{figure}
    \centering
    \includegraphics[width=1.0\linewidth]{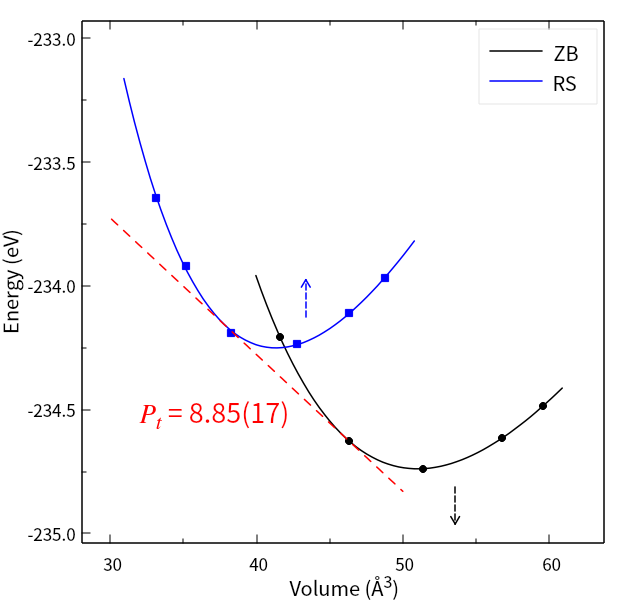}
    \caption{Equation of state curves for ZB (black solid lines) and RS (blue solid lines) phases of \ce{InP} (fitted to the third-order Birch-Murnaghan equation). Filled circles (ZB) and squares (RS) correspond to the calculated data from phaseless pw-AFQMC, with the associated stochastic errors being smaller than the marker size. All values, obtained using cubic supercell, are converted to the one-formula primitive cell for consistency in presentation. The red dashed line denotes the common tangent to the two EOS curves and its (negative) slope represents the transition pressure $P_t$, where the value in parentheses denotes the statistical error during fittings. Arrows indicate the direction of residual finite-size errors, as discussed in the main text.
    }
    \label{Eoss} 
\end{figure}

Predicting the transition pressure of materials constitutes a stringent benchmark for theoretical total-energy methods, as it requires high precision in the calculation of relative energies. 
For prototypical \Rmnum{3}-\Rmnum{5} compounds, the structure-property transformation (e.g. semiconductor to metal) with substantial volume collapse that accompanies pressure-induced transitions has been extensively investigated \cite{4_35,InP}. Here we focus on indium phosphide (\ce{InP}), a \Rmnum{3}-\Rmnum{5} compound containing heavy element \ce{In}, and employ our developed phaseless pw-AFQMC with SOC method to investigate its structural phases. Solid \ce{InP} serves as an ideal model system owing to its pressure-driven phase transition from the ZB to the RS phase with both structures possessing only a single structural variable parameter (the lattice constant), enabling comprehensive characterization of the transition via our phaseless pw-AFQMC calculations.

For both DFT and phaseless pw-AFQMC calculations, the FR pseudopotentials of \ce{In} and \ce{P} are generated using generalized gradient approximation (GGA) with the Perdew-Burke-Ernzerhof (PBE) \cite{PBE} exchange-correlation functional. A ``large-core'' pseudopotential retaining the $5s^25p^1$ valence states is adopted for \ce{In}, whereas \ce{P} used a neon-core pseudopotential. In phaseless pw-AFQMC calculations for both structures, a four-formula cubic supercell is used with the twist boundary condition corresponding to the Baldereschi $\mathbf{k}$ point. The fitted EOSs curves via the third-order Birch-Murnaghan equation \cite{BM} are presented in Fig.~\ref{Eoss}. The equilibrium lattice constant and the bulk modulus obtained from our phaseless pw-AFQMC and corresponding DFT calculations, alongside results from other theoretical and experimental works for comparison, are summarized in Table~\ref{tab:table1}. Taking the ZB phase as an illustration, it can be seen that standard GGA calculations systematically overestimate the equilibrium lattice constant, while LDA gives a slight underestimation. For the bulk modulus, the calculated value by LDA is consistent with the experiment, whereas those calculated by GGA are substantially underestimated. Notably, despite LDA’s accurate prediction of structural parameters, the phase transition pressure computed via LDA is more than $3$ GPa lower than the experimental value. By contrast, the transition pressure obtained from GGA calculations is in better agreement with experimental measurements than that from LDA. Compared with these theoretical and experimental results, the phaseless pw-AFQMC calculations yield high-accuracy structural properties for the (ZB) phase---in closer accord with LDA results---while delivering a far more reliable prediction of the phase transition pressure that is comparable to the GGA result. We also provide a qualitative analysis of the underlying origin for this behavior: LDA’s accurate estimation of structural parameters (assumed to hold for the RS phase as well) yet poor prediction of the transition pressure is most possibly attributed to an inherent bias in its evaluation of the relative total energy between the two phases.

The visible discrepancy between the phaseless pw-AFQMC-predicted phase transition pressure and experimental measurements can be attributed to the simplified computational treatments employed in this work. First and foremost, residual finite-size errors persist for both phases and exhibit opposing trends in their respective effects. We additionally repeated calculations using a one-formula primitive fcc cell, and the predicted structural parameters are reasonably satisfactory but with a low transition pressure of only $\sim7.2$ GPa. A comparison of the EOS curves for cubic versus primitive cells across both phases reveals that the larger supercell elevates the overall EOS curve for the RS phase while lowering that of the ZB phase, as indicated by the arrows in Fig.~\ref{Eoss}, consequently increasing and improving the predicted transition pressure. Second, since \ce{InP} is metallic in the RS phase, a denser sampling of random twist-angles would be required to achieve a rigorous, high-accuracy treatment of this phase. Furthermore, incorporating the semicore $4d$ states into the \ce{In} pseudopotential may potentially yield further improvements, as indicated by the slight discrepancy compared to those from referenced all-electron calculations (third row in Table~\ref{tab:table1}) in DFT level.

\section{SUMMARY} \label{sec:4}

We have explicitly incorporated SOC into the many-body phaseless pw-AFQMC method, and this implementation extends the capabilities of our approach to enable more accurate descriptions of heavy-element-containing systems beyond the SR approximation. This advancement is achieved by expanding the single-particle basis in a complete set of two-component spinor plane-wave basis and employing the FR ONCV pseudopotentials that inherently account for SOC effects. We discussed the formalisms within this framework, and outlined the implementation of phaseless pw-AFQMC with and without SOC. To demonstrate the effects of SOC, we apply the phaseless pw-AFQMC method to calculate the dissociation energy of molecule \ce{I_2} and the cohesive energy of bulk \ce{Pb} at both the SR and FR levels. Our calculated results are in good agreement with experimental measurements. Finally, we computed the structural properties of the \Rmnum{3}-\Rmnum{5} compound \ce{InP} in its ZB and RS phase. Our predicted structural properties for the ZB phase are in excellent agreement with experimental values, while the calculated phase transition pressure between the two phases is slightly beneath the experimental result. This deviation is primarily attributed to the distinct finite-size scaling behaviors of the two structures, for which recent advances, e.g., to impose ${\mathbf k}$-symmetry, improve computational efficiency, and implement GPU accelerations, can be easily incorporated.

\begin{acknowledgments}

This work was financially supported by the National Natural Science Foundation of China (Grants No. 12074040 and No. 11674027). F. Ma was also supported by the BNU Tang Scholar. The Flatiron Institute is a division of the Simons Foundation. Computing was carried out through the support from the Center for Advanced Quantum Studies, Beijing Normal University.
\end{acknowledgments}


\bibliography{Ref}

@PREAMBLE{
 "\providecommand{\noopsort}[1]{}" 
 # "\providecommand{\singleletter}[1]{#1}%" 
}

@article{SR,
doi = {10.1088/0022-3719/10/16/019},
url = {https://doi.org/10.1088/0022-3719/10/16/019},
year = {1977},
month = {aug},
publisher = {},
volume = {10},
number = {16},
pages = {3107},
author = {D D Koelling and B N Harmon},
_title = {A technique for relativistic spin-polarised calculations},
journal = {Journal of Physics C: Solid State Physics},
abstract = {A technique for reduction of the Dirac equation, which initially omits the spin-orbit interaction (thus keeping spin as a good quantum number), but retains all other relativistic kinematic effects such as mass-velocity, Darwin, and higher order terms is presented. The spin-orbit interaction can be included as a perturbation once the 'relativistic' spin-polarised bands and wavefunctions have been obtained. The technique is used together with the local spin density approximation for exchange and correlation to calculate the self-consistent charge and spin density of a neutral Gd atom. The calculated magnetic form factor agrees extremely well with experiment. Comparison with a paramagnetic RAPW calculation shows the procedure should be accurate and fast for general band structure determinations.}
}

@article{Kleinman_FR,
  _title = {Relativistic norm-conserving pseudopotential},
  author = {Kleinman, Leonard},
  journal = {Phys. Rev. B},
  volume = {21},
  issue = {6},
  pages = {2630--2631},
  numpages = {0},
  year = {1980},
  month = {Mar},
  publisher = {American Physical Society},
  doi = {10.1103/PhysRevB.21.2630},
  url = {https://link.aps.org/doi/10.1103/PhysRevB.21.2630}
}

@article{FR,
  _title = {Relativistic norm-conserving pseudopotentials},
  author = {Bachelet, Giovanni B. and Schl\"uter, M.},
  journal = {Phys. Rev. B},
  volume = {25},
  issue = {4},
  pages = {2103--2108},
  numpages = {0},
  year = {1982},
  month = {Feb},
  publisher = {American Physical Society},
  doi = {10.1103/PhysRevB.25.2103},
  url = {https://link.aps.org/doi/10.1103/PhysRevB.25.2103}
}

@article{USPP_SOC,
  _title = {Spin-orbit coupling with ultrasoft pseudopotentials: Application to Au and Pt},
  author = {Corso, Andrea Dal and Conte, Adriano Mosca},
  journal = {Phys. Rev. B},
  volume = {71},
  issue = {11},
  pages = {115106},
  numpages = {8},
  year = {2005},
  month = {Mar},
  publisher = {American Physical Society},
  doi = {10.1103/PhysRevB.71.115106},
  url = {https://link.aps.org/doi/10.1103/PhysRevB.71.115106}
}

@article{NCPP_SOC,
  _title = {Self-consistent treatment of spin-orbit coupling in solids using relativistic fully separable ab initio pseudopotentials},
  author = {Theurich, Gerhard and Hill, Nicola A.},
  journal = {Phys. Rev. B},
  volume = {64},
  issue = {7},
  pages = {073106},
  numpages = {4},
  year = {2001},
  month = {Jul},
  publisher = {American Physical Society},
  doi = {10.1103/PhysRevB.64.073106},
  url = {https://link.aps.org/doi/10.1103/PhysRevB.64.073106}
}

@article{NCPP_SRSO,
  _title = {First-principles calculations of spin-orbit splittings in solids using nonlocal separable pseudopotentials},
  author = {Hemstreet, L. A. and Fong, C. Y. and Nelson, J. S.},
  journal = {Phys. Rev. B},
  volume = {47},
  issue = {8},
  pages = {4238--4243},
  numpages = {0},
  year = {1993},
  month = {Feb},
  publisher = {American Physical Society},
  doi = {10.1103/PhysRevB.47.4238},
  url = {https://link.aps.org/doi/10.1103/PhysRevB.47.4238}
}

@article{DFT_Hohen,
  _title = {Inhomogeneous Electron Gas},
  author = {Hohenberg, P. and Kohn, W.},
  journal = {Phys. Rev.},
  volume = {136},
  issue = {3B},
  pages = {B864--B871},
  numpages = {0},
  year = {1964},
  month = {Nov},
  publisher = {American Physical Society},
  doi = {10.1103/PhysRev.136.B864},
  url = {https://link.aps.org/doi/10.1103/PhysRev.136.B864}
}

@article{DFT_KS,
  _title = {Self-Consistent Equations Including Exchange and Correlation Effects},
  author = {Kohn, W. and Sham, L. J.},
  journal = {Phys. Rev.},
  volume = {140},
  issue = {4A},
  pages = {A1133--A1138},
  numpages = {0},
  year = {1965},
  month = {Nov},
  publisher = {American Physical Society},
  doi = {10.1103/PhysRev.140.A1133},
  url = {https://link.aps.org/doi/10.1103/PhysRev.140.A1133}
}

@article{GTO_AFQMC1,
    author = {Al-Saidi, W. A. and Zhang, Shiwei and Krakauer, Henry},
    _title = {Auxiliary-field quantum Monte Carlo calculations of molecular systems with a Gaussian basis},
    journal = {The Journal of Chemical Physics},
    volume = {124},
    number = {22},
    pages = {224101},
    year = {2006},
    month = {06},
    abstract = {We extend the recently introduced phaseless auxiliary-field quantum Monte Carlo (QMC) approach to any single-particle basis and apply it to molecular systems with Gaussian basis sets. QMC methods in general scale favorably with the system size as a low power. A QMC approach with auxiliary fields, in principle, allows an exact solution of the Schrödinger equation in the chosen basis. However, the well-known sign/phase problem causes the statistical noise to increase exponentially. The phaseless method controls this problem by constraining the paths in the auxiliary-field path integrals with an approximate phase condition that depends on a trial wave function. In the present calculations, the trial wave function is a single Slater determinant from a Hartree-Fock calculation. The calculated all-electron total energies show typical systematic errors of no more than a few millihartrees compared to exact results. At equilibrium geometries in the molecules we studied, this accuracy is roughly comparable to that of coupled cluster with single and double excitations and with noniterative triples [CCSD(T)]. For stretched bonds in H2O, our method exhibits a better overall accuracy and a more uniform behavior than CCSD(T).},
    issn = {0021-9606},
    doi = {10.1063/1.2200885},
    url = {https://doi.org/10.1063/1.2200885},
    _eprint = {https://pubs.aip.org/aip/jcp/article-pdf/doi/10.1063/1.2200885/15386139/224101_1_online.pdf},
}

@article{GTO_AFQMC2,
    author = {Purwanto, Wirawan and Krakauer, Henry and Virgus, Yudistira and Zhang, Shiwei},
    _title = {Assessing weak hydrogen binding on Ca+ centers: An accurate many-body study with large basis sets},
    journal = {The Journal of Chemical Physics},
    volume = {135},
    number = {16},
    pages = {164105},
    year = {2011},
    month = {10},
    abstract = {Weak H2 physisorption energies present a significant challenge to even the best correlated theoretical many-body methods. We use the phaseless auxiliary-field quantum Monte Carlo method to accurately predict the binding energy of Ca+– 4H2. Attention has recently focused on this model chemistry to test the reliability of electronic structure methods for H2 binding on dispersed alkaline earth metal centers. A modified Cholesky decomposition is implemented to realize the Hubbard-Stratonovich transformation efficiently with large Gaussian basis sets. We employ the largest correlation-consistent Gaussian type basis sets available, up to cc-pCV5Z for Ca, to accurately extrapolate to the complete basis limit. The calculated potential energy curve exhibits binding with a double-well structure.},
    issn = {0021-9606},
    doi = {10.1063/1.3654002},
    url = {https://doi.org/10.1063/1.3654002},
    _eprint = {https://pubs.aip.org/aip/jcp/article-pdf/doi/10.1063/1.3654002/15444035/164105_1_online.pdf},
}

@article{Hubbard,
  _title = {Constrained path Monte Carlo method for fermion ground states},
  author = {Zhang, Shiwei and Carlson, J. and Gubernatis, J. E.},
  journal = {Phys. Rev. B},
  volume = {55},
  issue = {12},
  pages = {7464--7477},
  numpages = {0},
  year = {1997},
  month = {Mar},
  publisher = {American Physical Society},
  doi = {10.1103/PhysRevB.55.7464},
  url = {https://link.aps.org/doi/10.1103/PhysRevB.55.7464}
}

@article{GTO_SOC,
    author = {Eskridge, Brandon and Krakauer, Henry and Shi, Hao and Zhang, Shiwei},
    _title = {Ab initio calculations in atoms, molecules, and solids, treating spin–orbit coupling and electron interaction on an equal footing},
    journal = {The Journal of Chemical Physics},
    volume = {156},
    number = {1},
    pages = {014107},
    year = {2022},
    month = {01},
    abstract = {We incorporate explicit, non-perturbative treatment of spin–orbit coupling into ab initio auxiliary-field quantum Monte Carlo (AFQMC) calculations. The approach allows a general computational framework for molecular and bulk systems in which material specificity, electron correlation, and spin–orbit coupling effects can be captured accurately and on an equal footing, with favorable computational scaling vs system size. We adopt relativistic effective-core potentials that have been obtained by fitting to fully relativistic data and that have demonstrated a high degree of reliability and transferability in molecular systems. This results in a two-component spin-coupled Hamiltonian, which is then treated by generalizing the ab initio AFQMC approach. We demonstrate the method by computing the electron affinity in Pb, the bond dissociation energy in Br2 and I2, and solid Bi.},
    issn = {0021-9606},
    doi = {10.1063/5.0075900},
    url = {https://doi.org/10.1063/5.0075900},
    _eprint = {https://pubs.aip.org/aip/jcp/article-pdf/doi/10.1063/5.0075900/16531722/014107_1_online.pdf},
}

@article{Model_SOC,
_title = {Accurate computations of Rashba spin-orbit coupling in interacting systems: From the Fermi gas to real materials},
journal = {Journal of Physics and Chemistry of Solids},
volume = {128},
pages = {161-168},
year = {2019},
_note = {Spin-Orbit Coupled Materials},
issn = {0022-3697},
doi = {https://doi.org/10.1016/j.jpcs.2017.12.026},
url = {https://www.sciencedirect.com/science/article/pii/S0022369717316554},
author = {Peter Rosenberg and Hao Shi and Shiwei Zhang},
abstract = {We describe the treatment of Rashba spin-orbit coupling (SOC) in interacting many-fermion systems within the auxiliary-field quantum Monte Carlo framework, and present a set of illustrative results. These include numerically exact calculations on the ground-state properties of the spin-balanced, attractive two-dimensional Fermi gas, as well as a study of a tight-binding Hamiltonian with repulsive interaction. These systems are formally connected via the Hubbard Hamiltonian plus SOC, but cover different physics ranging from superfluidity and triplet pairing to SOC in real materials in the presence of strong interactions in localized orbitals. We carry out detailed benchmark studies of the method in the latter case when an approximation is needed to control the sign problem for repulsive Coulomb interactions. The methods presented here provide an approach for predictive computations in materials to study the interplay of SOC and strong correlation.}
}

@article{KB,
  _title = {Efficacious Form for Model Pseudopotentials},
  author = {Kleinman, Leonard and Bylander, D. M.},
  journal = {Phys. Rev. Lett.},
  volume = {48},
  issue = {20},
  pages = {1425--1428},
  numpages = {0},
  year = {1982},
  month = {May},
  publisher = {American Physical Society},
  doi = {10.1103/PhysRevLett.48.1425},
  url = {https://link.aps.org/doi/10.1103/PhysRevLett.48.1425}
}

@article{Hamann,
  _title = {Optimized norm-conserving Vanderbilt pseudopotentials},
  author = {Hamann, D. R.},
  journal = {Phys. Rev. B},
  volume = {88},
  issue = {8},
  pages = {085117},
  numpages = {10},
  year = {2013},
  month = {Aug},
  publisher = {American Physical Society},
  doi = {10.1103/PhysRevB.88.085117},
  url = {https://link.aps.org/doi/10.1103/PhysRevB.88.085117}
}

@article{AFQMC_ONCV,
  _title = {Auxiliary-field quantum Monte Carlo calculations with multiple-projector pseudopotentials},
  author = {Ma, Fengjie and Zhang, Shiwei and Krakauer, Henry},
  journal = {Phys. Rev. B},
  volume = {95},
  issue = {16},
  pages = {165103},
  numpages = {9},
  year = {2017},
  month = {Apr},
  publisher = {American Physical Society},
  doi = {10.1103/PhysRevB.95.165103},
  url = {https://link.aps.org/doi/10.1103/PhysRevB.95.165103}
}

@article{BO_app,
author = {Born, M. and Oppenheimer, R.},
journal = {Annalen der Physik},
volume = {389},
number = {20},
pages = {457-484},
doi = {https://doi.org/10.1002/andp.19273892002},
year = {1927}
}

@article{Ion_ion,
  _title = {Theory of lattice-dynamical properties of solids: Application to Si and Ge},
  author = {Yin, M. T. and Cohen, Marvin L.},
  journal = {Phys. Rev. B},
  volume = {26},
  issue = {6},
  pages = {3259--3272},
  numpages = {0},
  year = {1982},
  month = {Sep},
  publisher = {American Physical Society},
  doi = {10.1103/PhysRevB.26.3259},
  url = {https://link.aps.org/doi/10.1103/PhysRevB.26.3259}
}

@article{Ewald,
  author = {Fraser, Louisa M. and Foulkes, W. M. C. and Rajagopal, G. and Needs, R. J. and Kenny, S. D. and Williamson, A. J.},
  journal = {Phys. Rev. B},
  volume = {53},
  issue = {4},
  pages = {1814--1832},
  numpages = {0},
  year = {1996},
  month = {Jan},
  publisher = {American Physical Society},
  doi = {10.1103/PhysRevB.53.1814},
  url = {https://link.aps.org/doi/10.1103/PhysRevB.53.1814}
}

@article{Ewald_ori,
author = {Ewald, P. P.},
_title = {Die Berechnung optischer und elektrostatischer Gitterpotentiale},
journal = {Annalen der Physik},
volume = {369},
number = {3},
pages = {253-287},
doi = {https://doi.org/10.1002/andp.19213690304},
url = {https://onlinelibrary.wiley.com/doi/abs/10.1002/andp.19213690304},
_eprint = {https://onlinelibrary.wiley.com/doi/pdf/10.1002/andp.19213690304},
year = {1921}
}

@misc{Martin,
author = {Martin, Richard},
year = {Cambridge University Press{,} 2020},
month = {08},
pages = {},
title = {Electronic Structure: Basic Theory and Practical Methods},
isbn = {9781108429900},
doi = {10.1017/9781108555586}
}

@article{Van_USPP,
  _title = {Soft self-consistent pseudopotentials in a generalized eigenvalue formalism},
  author = {Vanderbilt, David},
  journal = {Phys. Rev. B},
  volume = {41},
  issue = {11},
  pages = {7892--7895},
  numpages = {0},
  year = {1990},
  month = {Apr},
  publisher = {American Physical Society},
  doi = {10.1103/PhysRevB.41.7892},
  url = {https://link.aps.org/doi/10.1103/PhysRevB.41.7892}
}

@article{Proj,
  _title = {Constrained path Monte Carlo method for fermion ground states},
  author = {Zhang, Shiwei and Carlson, J. and Gubernatis, J. E.},
  journal = {Phys. Rev. B},
  volume = {55},
  issue = {12},
  pages = {7464--7477},
  numpages = {0},
  year = {1997},
  month = {Mar},
  publisher = {American Physical Society},
  doi = {10.1103/PhysRevB.55.7464},
  url = {https://link.aps.org/doi/10.1103/PhysRevB.55.7464}
}

@article {Trotter,
    AUTHOR = {Trotter, H. F.},
     _title = {On the product of semi-groups of operators},
   JOURNAL = {Proc. Amer. Math. Soc.},
  FJOURNAL = {Proceedings of the American Mathematical Society},
    VOLUME = {10},
      YEAR = {1959},
     PAGES = {545--551},
      ISSN = {0002-9939,1088-6826},
   MRCLASS = {46.00},
  MRNUMBER = {108732},
MRREVIEWER = {G.\ Hufford},
       DOI = {10.2307/2033649},
       URL = {https://doi.org/10.2307/2033649},
}

@article{Suzuki,
  author = {Suzuki, Masuo},
  _title={Generalized Trotter's formula and systematic approximants of exponential operators and inner derivations with applications to many-body problems},
  journal = {Communications in Mathematical Physics},
  year = {1976},
  month = {Jun},
  day = {01},
  volume = {51},
  number={2},
  pages = {183--190},
  issn={1432-0916},
  doi = {10.1007/BF01609348},
  url = {https://doi.org/10.1007/BF01609348}
}

@article{Thouless1,
_title = {Stability conditions and nuclear rotations in the Hartree-Fock theory},
journal = {Nuclear Physics},
volume = {21},
pages = {225-232},
year = {1960},
issn = {0029-5582},
doi = {https://doi.org/10.1016/0029-5582(60)90048-1},
url = {https://www.sciencedirect.com/science/article/pii/0029558260900481},
author = {D.J. Thouless},
abstract = {An expression for a general Slater determinant is written in the notation of second quantization. This expression has just the right number of arbitrary coefficients, so no subsidiary conditions are required, and the expression for a particular determinant is unique. This notation is used to study two problems. Firstly, a condition for a particular solution of the Hartree-Fock equations to minimize the expectation value of the Hamiltonian is derived. This condition is equivalent to the condition for stability of collective modes in the random phase approximation. Secondly, the determinant which minimizes the expectation value of the Hamiltonian while giving a particular value to the expectation value of a component of angular momentum is found. In this way, an expression for the moment of inertia of an axially symmetric system is derived within the framework of the Hartree-Fock theory. The expression for a determinant is generalized to include the type of wave functions used in the theory of superconductivity.}
}

@article{Thouless2,
_title = {Vibrational states of nuclei in the random phase approximation},
journal = {Nuclear Physics},
volume = {22},
number = {1},
pages = {78-95},
year = {1961},
issn = {0029-5582},
doi = {https://doi.org/10.1016/0029-5582(61)90364-9},
url = {https://www.sciencedirect.com/science/article/pii/0029558261903649},
author = {D.J. Thouless},
abstract = {The properties of those states which can be obtained by exciting a single particle from the ground state of a nucleus are studied by using a perturbation-theory expansion for the Green's function. The formula obtained differs from the one used in a straightforward shall model calculation in some important respects. By comparing this formula with the condition for a solution of the Hartree-Fock equations to give a minimum of the expectation value of the energy, derived in an earlier paper, it is shown that it is always possible to choose an independent particle wave function which makes all collective modes stable in the random phase approximation. A study of the formal properties of solutions of the equation shows that the wave functions should be normalized using an indefinite metric: different eigenfunctions are then orthogonal. The eigenfunctions form a complete set, if there is no degeneracy in the solution of the Hartree-Fock equations, but not if there is a degeneracy. An upper bound for the lowest energy level is established, and the rule for calculating matrix elements of one-particle operators between the ground state and an excited state is given. It is shown to follow from the self-consistent field condition that the energy-weighted sum rules are preserved in this approximation. A discussion of spurious states is given, and it is shown that they separate out and have zero energy. A vector introduced in our earlier discussion of the cranking model is shown to be independent of all eigenvectors in this case.}
}

@article{Thouless_Ham,
  _title = {Energy measurement in auxiliary-field many-electron calculations},
  author = {Hamann, D. R. and Fahy, S. B.},
  journal = {Phys. Rev. B},
  volume = {41},
  issue = {16},
  pages = {11352--11363},
  numpages = {0},
  year = {1990},
  month = {Jun},
  publisher = {American Physical Society},
  doi = {10.1103/PhysRevB.41.11352},
  url = {https://link.aps.org/doi/10.1103/PhysRevB.41.11352}
}

@article{HS1,
  _title = {Calculation of Partition Functions},
  author = {Hubbard, J.},
  journal = {Phys. Rev. Lett.},
  volume = {3},
  issue = {2},
  pages = {77--78},
  numpages = {0},
  year = {1959},
  month = {Jul},
  publisher = {American Physical Society},
  doi = {10.1103/PhysRevLett.3.77},
  url = {https://link.aps.org/doi/10.1103/PhysRevLett.3.77}
}

@article{HS2,
author = {R. L. Stratonovich},
journal = {Dokl. Akad. Nauk SSSR},
_title = {A method for the. computation of quantum distribution functions},
year = {1957},
volume = {115},
issue = {6},
pages = {1097--1100},
}

@article{Phase_AFQMC,
  _title = {Quantum Monte Carlo Method using Phase-Free Random Walks with Slater Determinants},
  author = {Zhang, Shiwei and Krakauer, Henry},
  journal = {Phys. Rev. Lett.},
  volume = {90},
  issue = {13},
  pages = {136401},
  numpages = {4},
  year = {2003},
  month = {Apr},
  publisher = {American Physical Society},
  doi = {10.1103/PhysRevLett.90.136401},
  url = {https://link.aps.org/doi/10.1103/PhysRevLett.90.136401}
}

@article{AFQMC_mol,
  _title = {Phaseless auxiliary-field quantum Monte Carlo calculations with plane waves and pseudopotentials: Applications to atoms and molecules},
  author = {Suewattana, Malliga and Purwanto, Wirawan and Zhang, Shiwei and Krakauer, Henry and Walter, Eric J.},
  journal = {Phys. Rev. B},
  volume = {75},
  issue = {24},
  pages = {245123},
  numpages = {12},
  year = {2007},
  month = {Jun},
  publisher = {American Physical Society},
  doi = {10.1103/PhysRevB.75.245123},
  url = {https://link.aps.org/doi/10.1103/PhysRevB.75.245123}
}

@article{AFQMC_Green,
  _title = {Constrained path Monte Carlo method for fermion ground states},
  author = {Zhang, Shiwei and Carlson, J. and Gubernatis, J. E.},
  journal = {Phys. Rev. B},
  volume = {55},
  issue = {12},
  pages = {7464--7477},
  numpages = {0},
  year = {1997},
  month = {Mar},
  publisher = {American Physical Society},
  doi = {10.1103/PhysRevB.55.7464},
  url = {https://link.aps.org/doi/10.1103/PhysRevB.55.7464}
}

@article{AFQMC_BP,
  _title = {Quantum Monte Carlo method for the ground state of many-boson systems},
  author = {Purwanto, Wirawan and Zhang, Shiwei},
  journal = {Phys. Rev. E},
  volume = {70},
  issue = {5},
  pages = {056702},
  numpages = {18},
  year = {2004},
  month = {Nov},
  publisher = {American Physical Society},
  doi = {10.1103/PhysRevE.70.056702},
  url = {https://link.aps.org/doi/10.1103/PhysRevE.70.056702}
}

@article{PWAFQMC_BP_Den,
  _title = {Ab initio electronic density in solids by many-body plane-wave auxiliary-field quantum Monte Carlo calculations},
  author = {Chen, Siyuan and Motta, Mario and Ma, Fengjie and Zhang, Shiwei},
  journal = {Phys. Rev. B},
  volume = {103},
  issue = {7},
  pages = {075138},
  numpages = {13},
  year = {2021},
  month = {Feb},
  publisher = {American Physical Society},
  doi = {10.1103/PhysRevB.103.075138},
  url = {https://link.aps.org/doi/10.1103/PhysRevB.103.075138}
}

@article{Santos_Wick,
author = {dos Santos, Raimundo},
year = {2003},
month = {03},
pages = {},
_title = {Introduction to Quantum Monte Carlo simulations for fermionic systems},
volume = {33},
journal = {Brazilian Journal of Physics},
doi = {10.1590/S0103-97332003000100003}
}

@article{SOC_atom_mol,
    author = {Liu, Junzi and Cheng, Lan},
    _title = {An atomic mean-field spin-orbit approach within exact two-component theory for a non-perturbative treatment of spin-orbit coupling},
    journal = {The Journal of Chemical Physics},
    volume = {148},
    number = {14},
    pages = {144108},
    year = {2018},
    month = {04},
    abstract = {An atomic mean-field (AMF) spin-orbit (SO) approach within exact two-component theory (X2C) is reported, thereby exploiting the exact decoupling scheme of X2C, the one-electron approximation for the scalar-relativistic contributions, the mean-field approximation for the treatment of the two-electron SO contribution, and the local nature of the SO interactions. The Hamiltonian of the proposed SOX2CAMF scheme comprises the one-electron X2C Hamiltonian, the instantaneous two-electron Coulomb interaction, and an AMF SO term derived from spherically averaged Dirac-Coulomb Hartree-Fock calculations of atoms; no molecular relativistic two-electron integrals are required. Benchmark calculations for bond lengths, harmonic frequencies, dipole moments, and electric-field gradients for a set of diatomic molecules containing elements across the periodic table show that the SOX2CAMF scheme offers a balanced treatment for SO and scalar-relativistic effects and appears to be a promising candidate for applications to heavy-element containing systems. SOX2CAMF coupled-cluster calculations of molecular properties for bismuth compounds (BiN, BiP, BiF, BiCl, and BiI) are also presented and compared with experimental results to further demonstrate the accuracy and applicability of the SOX2CAMF scheme.},
    issn = {0021-9606},
    doi = {10.1063/1.5023750},
    url = {https://doi.org/10.1063/1.5023750},
    _eprint = {https://pubs.aip.org/aip/jcp/article-pdf/doi/10.1063/1.5023750/13549142/144108_1_online.pdf},
}

@article{SOC_element,
author = {Zhang, Boyi and Vandezande, Jonathon E. and Reynolds, Ryan D. and Schaefer, Henry F. III},
_title = {Spin–Orbit Coupling via Four-Component Multireference Methods: Benchmarking on p-Block Elements and Tentative Recommendations},
journal = {Journal of Chemical Theory and Computation},
volume = {14},
number = {3},
pages = {1235-1246},
year = {2018},
doi = {10.1021/acs.jctc.7b00989},
note ={PMID: 29461828},
URL = {https://doi.org/10.1021/acs.jctc.7b00989},
_eprint = {https://doi.org/10.1021/acs.jctc.7b00989}
}

@article{HalogenSOC,
    author = {Visscher, L. and Dyall, K. G.},
    _title = {Relativistic and correlation effects on molecular properties. I. The dihalogens F2, Cl2, Br2, I2, and At2},
    journal = {The Journal of Chemical Physics},
    volume = {104},
    number = {22},
    pages = {9040-9046},
    year = {1996},
    month = {06},
    abstract = {A benchmark study of a number of relativistic correlation methods is presented. Bond lengths, harmonic frequencies, and dissociation energies of the molecules F2, Cl2, Br2, I2, and At2 are calculated at various levels of theory, using both the Schrödinger and the Dirac–Coulomb–(Gaunt) Hamiltonian.},
    issn = {0021-9606},
    doi = {10.1063/1.471636},
    url = {https://doi.org/10.1063/1.471636},
    _eprint = {https://pubs.aip.org/aip/jcp/article-pdf/104/22/9040/19295916/9040_1_online.pdf},
}

@article{QE1,
doi = {10.1088/0953-8984/21/39/395502},
url = {https://doi.org/10.1088/0953-8984/21/39/395502},
year = {2009},
month = {sep},
publisher = {},
volume = {21},
number = {39},
pages = {395502},
author = {Giannozzi, Paolo and Baroni, Stefano and Bonini, Nicola and Calandra, Matteo and Car, Roberto and Cavazzoni, Carlo and Ceresoli, Davide and Chiarotti, Guido L and Cococcioni, Matteo and Dabo, Ismaila and Dal Corso, Andrea and de Gironcoli, Stefano and Fabris, Stefano and Fratesi, Guido and Gebauer, Ralph and Gerstmann, Uwe and Gougoussis, Christos and Kokalj, Anton and Lazzeri, Michele and Martin-Samos, Layla and Marzari, Nicola and Mauri, Francesco and Mazzarello, Riccardo and Paolini, Stefano and Pasquarello, Alfredo and Paulatto, Lorenzo and Sbraccia, Carlo and Scandolo, Sandro and Sclauzero, Gabriele and Seitsonen, Ari P and Smogunov, Alexander and Umari, Paolo and Wentzcovitch, Renata M},
_title = {QUANTUM ESPRESSO: a modular and open-source software project for quantum
simulations of materials},
journal = {Journal of Physics: Condensed Matter},
abstract = {QUANTUM ESPRESSO is an integrated suite of computer codes for electronic-structure calculations and materials modeling, based on density-functional theory, plane waves, and pseudopotentials (norm-conserving, ultrasoft, and projector-augmented wave). The acronym ESPRESSO stands for opEn Source Package for Research in Electronic Structure, Simulation, and Optimization. It is freely available to researchers around the world under the terms of the GNU General Public License. QUANTUM ESPRESSO builds upon newly-restructured electronic-structure codes that have been developed and tested by some of the original authors of novel electronic-structure algorithms and applied in the last twenty years by some of the leading materials modeling groups worldwide. Innovation and efficiency are still its main focus, with special attention paid to massively parallel architectures, and a great effort being devoted to user friendliness. QUANTUM ESPRESSO is evolving towards a distribution of independent and interoperable codes in the spirit of an open-source project, where researchers active in the field of electronic-structure calculations are encouraged to participate in the project by contributing their own codes or by implementing their own ideas into existing codes.}
}

@article{QE2,
doi = {10.1088/1361-648X/aa8f79},
url = {https://doi.org/10.1088/1361-648X/aa8f79},
year = {2017},
month = {oct},
publisher = {IOP Publishing},
volume = {29},
number = {46},
pages = {465901},
author = {Giannozzi, P and Andreussi, O and Brumme, T and Bunau, O and Buongiorno Nardelli, M and Calandra, M and Car, R and Cavazzoni, C and Ceresoli, D and Cococcioni, M and Colonna, N and Carnimeo, I and Dal Corso, A and de Gironcoli, S and Delugas, P and DiStasio, R A and Ferretti, A and Floris, A and Fratesi, G and Fugallo, G and Gebauer, R and Gerstmann, U and Giustino, F and Gorni, T and Jia, J and Kawamura, M and Ko, H-Y and Kokalj, A and Küçükbenli, E and Lazzeri, M and Marsili, M and Marzari, N and Mauri, F and Nguyen, N L and Nguyen, H-V and Otero-de-la-Roza, A and Paulatto, L and Poncé, S and Rocca, D and Sabatini, R and Santra, B and Schlipf, M and Seitsonen, A P and Smogunov, A and Timrov, I and Thonhauser, T and Umari, P and Vast, N and Wu, X and Baroni, S},
_title = {Advanced capabilities for materials modelling with Quantum ESPRESSO},
journal = {Journal of Physics: Condensed Matter},
abstract = {Quantum ESPRESSO is an integrated suite of open-source computer codes for quantum simulations of materials using state-of-the-art electronic-structure techniques, based on density-functional theory, density-functional perturbation theory, and many-body perturbation theory, within the plane-wave pseudopotential and projector-augmented-wave approaches. Quantum ESPRESSO owes its popularity to the wide variety of properties and processes it allows to simulate, to its performance on an increasingly broad array of hardware architectures, and to a community of researchers that rely on its capabilities as a core open-source development platform to implement their ideas. In this paper we describe recent extensions and improvements, covering new methodologies and property calculators, improved parallelization, code modularization, and extended interoperability both within the distribution and with external software.}
}

@article{Abinit,
_title = {First-principles computation of material properties: the ABINIT software project},
journal = {Computational Materials Science},
volume = {25},
number = {3},
pages = {478-492},
year = {2002},
issn = {0927-0256},
doi = {https://doi.org/10.1016/S0927-0256(02)00325-7},
url = {https://www.sciencedirect.com/science/article/pii/S0927025602003257},
author = {X. Gonze and J.-M. Beuken and R. Caracas and F. Detraux and M. Fuchs and G.-M. Rignanese and L. Sindic and M. Verstraete and G. Zerah and F. Jollet and M. Torrent and A. Roy and M. Mikami and Ph. Ghosez and J.-Y. Raty and D.C. Allan},
keywords = {Density functional theory, Software engineering, Electronic structure},
abstract = {The density functional theory (DFT) computation of electronic structure, total energy and other properties of materials, is a field in constant progress. In order to stay at the forefront of knowledge, a DFT software project can benefit enormously from widespread collaboration, if handled properly. Also, modern software engineering concepts can considerably ease its development. The ABINIT project relies upon these ideas: freedom of sources, reliability, portability, and self-documentation are emphasised, in the development of a sophisticated plane-wave pseudopotential code. We describe ABINITv3.0, distributed under the GNU General Public License. The list of ABINITv3.0 capabilities is presented, as well as the different software techniques that have been used until now: PERL scripts and CPP directives treat a unique set of FORTRAN90 source files to generate sequential (or parallel) object code for many different platforms; more than 200 automated tests secure existing capabilities; strict coding rules are followed; the documentation is extensive, including online help files, tutorials, and HTML-formatted sources.}
}

@misc{I2_De,
 _title={Molecular Spectra and Molecular Structure, IV. Constants of Diatomic Molecules},
  author={Huber, K. P. and Herzberg, Gerhard},
  publisher={Springer New York, NY},
  year={Springer New York{,} NY{,} 1979},
  doi = {10.1007/978-1-4757-0961-2},
}

@misc{ONCVPSP,
 author = {D. R. Hamann},
_title = {Open-source pseudopotential code ONCVPSP},
 howpublished = {\url{http://www.mat-simresearch.com/}},
}

@article{SG15,
_title = {Optimization algorithm for the generation of ONCV pseudopotentials},
journal = {Computer Physics Communications},
volume = {196},
pages = {36-44},
year = {2015},
issn = {0010-4655},
doi = {https://doi.org/10.1016/j.cpc.2015.05.011},
url = {https://www.sciencedirect.com/science/article/pii/S0010465515001897},
author = {Martin Schlipf and François Gygi},
keywords = {Density functional theory, Pseudopotential, Plane wave, All-electron calculation, Condensed matter},
abstract = {We present an optimization algorithm to construct pseudopotentials and use it to generate a set of Optimized Norm-Conserving Vanderbilt (ONCV) pseudopotentials for elements up to Z=83 (Bi) (excluding Lanthanides). We introduce a quality function that assesses the agreement of a pseudopotential calculation with all-electron FLAPW results, and the necessary plane-wave energy cutoff. This quality function allows us to use a Nelder–Mead optimization algorithm on a training set of materials to optimize the input parameters of the pseudopotential construction for most of the periodic table. We control the accuracy of the resulting pseudopotentials on a test set of materials independent of the training set. We find that the automatically constructed pseudopotentials (http://www.quantum-simulation.org) provide a good agreement with the all-electron results obtained using the FLEUR code with a plane-wave energy cutoff of approximately 60 Ry.}
}

@article{FS_corr,
  _title = {Finite-Size Correction in Many-Body Electronic Structure Calculations},
  author = {Kwee, Hendra and Zhang, Shiwei and Krakauer, Henry},
  journal = {Phys. Rev. Lett.},
  volume = {100},
  issue = {12},
  pages = {126404},
  numpages = {4},
  year = {2008},
  month = {Mar},
  publisher = {American Physical Society},
  doi = {10.1103/PhysRevLett.100.126404},
  url = {https://link.aps.org/doi/10.1103/PhysRevLett.100.126404}
}

@article{FS_mag_corr,
  _title = {Finite-size correction in many-body electronic structure calculations of magnetic systems},
  author = {Ma, Fengjie and Zhang, Shiwei and Krakauer, Henry},
  journal = {Phys. Rev. B},
  volume = {84},
  issue = {15},
  pages = {155130},
  numpages = {9},
  year = {2011},
  month = {Oct},
  publisher = {American Physical Society},
  doi = {10.1103/PhysRevB.84.155130},
  url = {https://link.aps.org/doi/10.1103/PhysRevB.84.155130}
}

@article{I2_SR,
    author = {Peterson, Kirk A. and Figgen, Detlev and Goll, Erich and Stoll, Hermann and Dolg, Michael},
    _title = {Systematically convergent basis sets with relativistic pseudopotentials. II. Small-core pseudopotentials and correlation consistent basis sets for the post-d group 16–18 elements},
    journal = {The Journal of Chemical Physics},
    volume = {119},
    number = {21},
    pages = {11113-11123},
    year = {2003},
    month = {12},
    abstract = {A series of correlation consistent basis sets have been developed for the post-d group 16–18 elements in conjunction with small-core relativistic pseudopotentials of the energy-consistent variety. The latter were adjusted to multiconfiguration Dirac–Hartree–Fock data based on the Dirac–Coulomb–Breit Hamiltonian. The outer-core (n−1)spd shells are explicitly treated together with the nsp valence shell with these PPs. The accompanying cc-pVnZ-PP and aug-cc-pVnZ-PP basis sets range in size from DZ to 5Z quality and yield systematic convergence of both Hartree–Fock and correlated total energies. In addition to the calculation of atomic electron affinities and dipole polarizabilities of the rare gas atoms, numerous molecular benchmark calculations (HBr, HI, HAt, Br2, I2, At2, SiSe, SiTe, SiPo, KrH+, XeH+, and RnH+) are also reported at the coupled cluster level of theory. For the purposes of comparison, all-electron calculations using the Douglas–Kroll–Hess Hamiltonian have also been carried out for the halogen-containing molecules using basis sets of 5Z quality.},
    issn = {0021-9606},
    doi = {10.1063/1.1622924},
    url = {https://doi.org/10.1063/1.1622924},
    _eprint = {https://pubs.aip.org/aip/jcp/article-pdf/119/21/11113/19321245/11113_1_online.pdf},
}

@misc{Pb_coh,
    author = {L. Brewer},
   _title = {THE COHESIVE ENERGIES OF THE ELEMENTS},
    institution = {Lawrence Berkeley National Laboratory},
    year = {Lawrence Berkeley National Laboratory{,} 1977},
    number = {LBL-3720 Rev.},
    url = {https://escholarship.org/uc/item/08p2578m},
}

@article{AFQMC_press,
  _title = {Pressure-induced diamond to $\ensuremath{\beta}$-tin transition in bulk silicon: A quantum Monte Carlo study},
  author = {Purwanto, Wirawan and Krakauer, Henry and Zhang, Shiwei},
  journal = {Phys. Rev. B},
  volume = {80},
  issue = {21},
  pages = {214116},
  numpages = {9},
  year = {2009},
  month = {Dec},
  publisher = {American Physical Society},
  doi = {10.1103/PhysRevB.80.214116},
  url = {https://link.aps.org/doi/10.1103/PhysRevB.80.214116}
}

@article{MP,
  _title = {Special points for Brillouin-zone integrations},
  author = {Monkhorst, Hendrik J. and Pack, James D.},
  journal = {Phys. Rev. B},
  volume = {13},
  issue = {12},
  pages = {5188--5192},
  numpages = {0},
  year = {1976},
  month = {Jun},
  publisher = {American Physical Society},
  doi = {10.1103/PhysRevB.13.5188},
  url = {https://link.aps.org/doi/10.1103/PhysRevB.13.5188}
}

@article{Bald,
  _title = {Mean-Value Point in the Brillouin Zone},
  author = {Baldereschi, A.},
  journal = {Phys. Rev. B},
  volume = {7},
  issue = {12},
  pages = {5212--5215},
  numpages = {0},
  year = {1973},
  month = {Jun},
  publisher = {American Physical Society},
  doi = {10.1103/PhysRevB.7.5212},
  url = {https://link.aps.org/doi/10.1103/PhysRevB.7.5212}
}

@article{4_35,
  _title = {High-pressure phases of group-IV, III--V, and II--VI compounds},
  author = {Mujica, A. and Rubio, Angel and Mu\~noz, A. and Needs, R. J.},
  journal = {Rev. Mod. Phys.},
  volume = {75},
  issue = {3},
  pages = {863--912},
  numpages = {0},
  year = {2003},
  month = {Jul},
  publisher = {American Physical Society},
  doi = {10.1103/RevModPhys.75.863},
  url = {https://link.aps.org/doi/10.1103/RevModPhys.75.863}
}

@article{InP,
  _title = {High-pressure phases of III-V zinc-blende semiconductors},
  author = {Zhang, S. B. and Cohen, Marvin L.},
  journal = {Phys. Rev. B},
  volume = {35},
  issue = {14},
  pages = {7604--7610},
  numpages = {0},
  year = {1987},
  month = {May},
  publisher = {American Physical Society},
  doi = {10.1103/PhysRevB.35.7604},
  url = {https://link.aps.org/doi/10.1103/PhysRevB.35.7604}
}

@article{InP_Pt,
  _title = {Equation of state of InP to 19 GPa},
  author = {Menoni, Carmen S. and Spain, Ian L.},
  journal = {Phys. Rev. B},
  volume = {35},
  issue = {14},
  pages = {7520--7525},
  numpages = {0},
  year = {1987},
  month = {May},
  publisher = {American Physical Society},
  doi = {10.1103/PhysRevB.35.7520},
  url = {https://link.aps.org/doi/10.1103/PhysRevB.35.7520}
}

@article{PBE,
  _title = {Generalized Gradient Approximation Made Simple},
  author = {Perdew, John P. and Burke, Kieron and Ernzerhof, Matthias},
  journal = {Phys. Rev. Lett.},
  volume = {77},
  issue = {18},
  pages = {3865--3868},
  numpages = {0},
  year = {1996},
  month = {Oct},
  publisher = {American Physical Society},
  doi = {10.1103/PhysRevLett.77.3865},
  url = {https://link.aps.org/doi/10.1103/PhysRevLett.77.3865}
}

@article{BM,
doi={10.1073/pnas.30.9.244},
author = {F. D. Murnaghan },
_title = {The Compressibility of Media under Extreme Pressures},
journal = {Proceedings of the National Academy of Sciences},
volume = {30},
number = {9},
pages = {244-247},
year = {1944},
doi = {10.1073/pnas.30.9.244},
URL = {https://www.pnas.org/doi/abs/10.1073/pnas.30.9.244},
_eprint = {https://www.pnas.org/doi/pdf/10.1073/pnas.30.9.244}
}

@article{HSE,
    author = {Schimka, Laurids and Harl, Judith and Kresse, Georg},
    _title = {Improved hybrid functional for solids: The HSEsol functional},
    journal = {The Journal of Chemical Physics},
    volume = {134},
    number = {2},
    pages = {024116},
    year = {2011},
    month = {01},
    abstract = {We introduce the hybrid functional HSEsol. It is based on PBEsol, a revised Perdew–Burke–Ernzerhof functional, designed to yield accurate equilibrium properties for solids and their surfaces. We present lattice constants, bulk moduli, atomization energies, heats of formation, and band gaps for extended systems, as well as atomization energies for the molecular G2-1 test set. Compared to HSE, significant improvements are found for lattice constants and atomization energies of solids, but atomization energies of molecules are slightly worse than for HSE. Additionally, we present zero-point anharmonic expansion corrections to the lattice constants and bulk moduli, evaluated from ab initio phonon calculations.},
    issn = {0021-9606},
    doi = {10.1063/1.3524336},
    url = {https://doi.org/10.1063/1.3524336},
    _eprint = {https://pubs.aip.org/aip/jcp/article-pdf/doi/10.1063/1.3524336/13040292/024116_1_online.pdf},
}

@misc{Cry_Wyckoff,
    author = {R. W. G. Wyckoff},
   _title = {Crystal structures, Second edition},
    publisher = {Wiley} ,
    year = {Wiley{,} 1963},
    doi = {10.1107/S0365110X65000361}
}

@article{InP_Pt_DFT1,
_title = {Structural, electronic and optical properties of InP under pressure: An ab-initio study},
journal = {Computational Condensed Matter},
volume = {17},
pages = {e00333},
year = {2018},
issn = {2352-2143},
doi = {https://doi.org/10.1016/j.cocom.2018.e00333},
url = {https://www.sciencedirect.com/science/article/pii/S2352214318301576},
author = {A. Baida and M. Ghezali},
keywords = {DFT, FP-LAPW, mBJ+LDA/GGA, InP, Pressure},
abstract = {In this paper, we presented an ab-initio study of the structural, electronic and optical properties of the binary compound of indium phosphide (InP). This material is widely used in the field of optoelectronics and microelectronics fast. Our calculations were made by the method of augmented plane wave (FP-LAPW), based on the density functional theory (DFT) and implemented in the calculation code Wien2k. This calculus of the electronic band, structure and optical properties were performed using local-density approximation (LDA), generalized gradient approximation (GGA), and a combination of modified Becke—Johnson exchange potential plus LDA and GGA (mBJ + LDA/GGA) for exchange–correlation potential. We find in our calculations, the usual trends, that the (GGA) unlike LDA overestimates the lattice parameter and underestimates the bulk modulus. The phase transitions for this material from structure B3 to structures Imm2, B8-1, B10 and B2 are possible under low pressures and The calculation of the density of state gives a detailed explanation of the contribution of the different orbitals.}
}

@article{InP_Pt_DFT2,
_title = {First-principles study on structural properties and phase stability of III-phosphide (BP, GaP, AlP and InP)},
journal = {Computational Materials Science},
volume = {47},
number = {3},
pages = {685-692},
year = {2010},
issn = {0927-0256},
doi = {https://doi.org/10.1016/j.commatsci.2009.10.009},
url = {https://www.sciencedirect.com/science/article/pii/S0927025609003930},
author = {O. Arbouche and B. Belgoumène and B. Soudini and Y. Azzaz and H. Bendaoud and K. Amara},
keywords = {III-Phosphide (BP, GaP, AlP, InP), Lattice parameter, Bulk modulus, Pressure, Phase transition, FP-LAPW+lo, GGA},
abstract = {We have performed first principles total-energy calculations based on the full-potential augmented plane-wave plus local orbitals (FP-LAPW+lo) method in order to investigate the phase transformations under high pressure of III-phosphide (BP, GaP, AlP, InP) in the zinc-blende, NaCl, CsCl, sc16, cmcm, d-β-tin, NiAs, Immm, and Imm2 structures. The generalized gradient approximation (GGA) was used for the exchange and correlation energy density functional. Our work relates directly to recent experimental work on III-phosphide compounds. The ground-state properties such as lattice parameter, bulk modulus and its pressure derivative as well as the structural phase stability. Along with previous work, we now have enough theoretical results to support a different systematics of the high-pressure phases of III-phosphide compounds.}
}

@article{RevModPhys.89.040502,
 _title = {Nobel Lecture: Topological quantum matter},
  author = {Haldane, F. Duncan M.},
  journal = {Rev. Mod. Phys.},
  volume = {89},
  issue = {4},
  pages = {040502},
  numpages = {10},
  year = {2017},
  month = {Oct},
  publisher = {American Physical Society},
  doi = {10.1103/RevModPhys.89.040502},
  url = {https://link.aps.org/doi/10.1103/RevModPhys.89.040502}
}

@article{RevModPhys.83.1057,
 _title = {Topological insulators and superconductors},
  author = {Qi, Xiao-Liang and Zhang, Shou-Cheng},
  journal = {Rev. Mod. Phys.},
  volume = {83},
  issue = {4},
  pages = {1057--1110},
  numpages = {0},
  year = {2011},
  month = {Oct},
  publisher = {American Physical Society},
  doi = {10.1103/RevModPhys.83.1057},
  url = {https://link.aps.org/doi/10.1103/RevModPhys.83.1057}
}

@article{RevModPhys.82.3045,
 _title = {Colloquium: Topological insulators},
  author = {Hasan, M. Z. and Kane, C. L.},
  journal = {Rev. Mod. Phys.},
  volume = {82},
  issue = {4},
  pages = {3045--3067},
  numpages = {0},
  year = {2010},
  month = {Nov},
  publisher = {American Physical Society},
  doi = {10.1103/RevModPhys.82.3045},
  url = {https://link.aps.org/doi/10.1103/RevModPhys.82.3045}
}

@article{RevModPhys.90.015001,
 _title = {Weyl and Dirac semimetals in three-dimensional solids},
  author = {Armitage, N. P. and Mele, E. J. and Vishwanath, Ashvin},
  journal = {Rev. Mod. Phys.},
  volume = {90},
  issue = {1},
  pages = {015001},
  numpages = {57},
  year = {2018},
  month = {Jan},
  publisher = {American Physical Society},
  doi = {10.1103/RevModPhys.90.015001},
  url = {https://link.aps.org/doi/10.1103/RevModPhys.90.015001}
}

@misc{Jackson,
author = {Jackson, John David},
year = {Wiley, Hoboken, New Jersey{,} 1998},
title = {Classical Electrodynamics, 3rd Ed.},
isbn = {9780471309321}
}

@misc{Tsymbal,
author = {Tsymbal, E. Y. and I. \v{Z}uti\'c (eds.)},
year = {CRC Press, Boca Raton, FL{,} 2019},
title = {Spintronics Handbook, 2nd Ed.},
isbn = {9780367777876}
}

@article{3m4m-3v59,
 _title = {Kitaev quantum spin liquids},
  author = {Matsuda, Yuji and Shibauchi, Takasada and Kee, Hae-Young},
  journal = {Rev. Mod. Phys.},
  volume = {97},
  issue = {4},
  pages = {045003},
  numpages = {61},
  year = {2025},
  month = {Dec},
  publisher = {American Physical Society},
  doi = {10.1103/3m4m-3v59},
  url = {https://link.aps.org/doi/10.1103/3m4m-3v59}
}

@article{RevModPhys.90.015005,
 _title = {Antiferromagnetic spintronics},
  author = {Baltz, V. and Manchon, A. and Tsoi, M. and Moriyama, T. and Ono, T. and Tserkovnyak, Y.},
  journal = {Rev. Mod. Phys.},
  volume = {90},
  issue = {1},
  pages = {015005},
  numpages = {57},
  year = {2018},
  month = {Feb},
  publisher = {American Physical Society},
  doi = {10.1103/RevModPhys.90.015005},
  url = {https://link.aps.org/doi/10.1103/RevModPhys.90.015005}
}

@article{RevModPhys.76.323,
 _title = {Spintronics: Fundamentals and applications},
  author = {\ifmmode \check{Z}\else \v{Z}\fi{}uti\ifmmode \acute{c}\else \'{c}\fi{}, Igor and Fabian, Jaroslav and Das Sarma, S.},
  journal = {Rev. Mod. Phys.},
  volume = {76},
  issue = {2},
  pages = {323--410},
  numpages = {0},
  year = {2004},
  month = {Apr},
  publisher = {American Physical Society},
  doi = {10.1103/RevModPhys.76.323},
  url = {https://link.aps.org/doi/10.1103/RevModPhys.76.323}
}

@article{RevModPhys.82.53,
 _title = {Magnetic pyrochlore oxides},
  author = {Gardner, Jason S. and Gingras, Michel J. P. and Greedan, John E.},
  journal = {Rev. Mod. Phys.},
  volume = {82},
  issue = {1},
  pages = {53--107},
  numpages = {0},
  year = {2010},
  month = {Jan},
  publisher = {American Physical Society},
  doi = {10.1103/RevModPhys.82.53},
  url = {https://link.aps.org/doi/10.1103/RevModPhys.82.53}
}

@article{RevModPhys.97.031001,
 _title = {Colloquium: Quantum properties and functionalities of magnetic skyrmions},
  author = {Petrovi\ifmmode \acute{c}\else \'{c}\fi{}, Alexander P. and Psaroudaki, Christina and Fischer, Peter and Garst, Markus and Panagopoulos, Christos},
  journal = {Rev. Mod. Phys.},
  volume = {97},
  issue = {3},
  pages = {031001},
  numpages = {31},
  year = {2025},
  month = {Jul},
  publisher = {American Physical Society},
  doi = {10.1103/RevModPhys.97.031001},
  url = {https://link.aps.org/doi/10.1103/RevModPhys.97.031001}
}

@article{RevModPhys.95.011002,
 _title = {Colloquium: Quantum anomalous Hall effect},
  author = {Chang, Cui-Zu and Liu, Chao-Xing and MacDonald, Allan H.},
  journal = {Rev. Mod. Phys.},
  volume = {95},
  issue = {1},
  pages = {011002},
  numpages = {33},
  year = {2023},
  month = {Jan},
  publisher = {American Physical Society},
  doi = {10.1103/RevModPhys.95.011002},
  url = {https://link.aps.org/doi/10.1103/RevModPhys.95.011002}
}

@article{PhysRevB.90.041112,
 _title = {$\ensuremath{\alpha}\ensuremath{-}{\mathrm{RuCl}}_{3}$: A spin-orbit assisted Mott insulator on a honeycomb lattice},
  author = {Plumb, K. W. and Clancy, J. P. and Sandilands, L. J. and Shankar, V. Vijay and Hu, Y. F. and Burch, K. S. and Kee, Hae-Young and Kim, Young-June},
  journal = {Phys. Rev. B},
  volume = {90},
  issue = {4},
  pages = {041112},
  numpages = {5},
  year = {2014},
  month = {Jul},
  publisher = {American Physical Society},
  doi = {10.1103/PhysRevB.90.041112},
  url = {https://link.aps.org/doi/10.1103/PhysRevB.90.041112}
}

@article{APR2025,
    author = {Banerjee, N. and Bell, C. and Ciccarelli, C. and Hesjedal, T. and Johnson, F. and Kurebayashi, H. and Moore, T. A. and Moutafis, C. and Stern, H. L. and Vera-Marun, I. J. and Wade, J. and Barton, C. and Connolly, M. R. and Curson, N. J. and Fallon, K. and Fisher, A. J. and Gangloff, D. A. and Griggs, W. and Linfield, E. and Marrows, C. H. and Rossi, A. and Schindler, F. and Smith, J. and Thomson, T. and Kazakova, O.},
   _title = {Materials for quantum technologies: A roadmap for spin and topology},
    journal = {Applied Physics Reviews},
    volume = {12},
    number = {4},
    pages = {041328},
    year = {2025},
    month = {12},
    abstract = {In this perspective article, we explore some of the promising spin and topology material platforms (e.g., spins in semiconductors and superconductors, skyrmionic, topological, and two-dimensional materials) being developed for such quantum components as qubits, superconducting memories, sensing, and metrological standards, and discuss their figures of merit. Spin- and topology-related quantum phenomena have several advantages, including high coherence time, topological protection and stability, low error rate, relative ease of engineering and control, and simple initiation and readout. However, the relevant technologies are at different stages of research and development, and here, we discuss their state-of-the-art, potential applications, challenges, and solutions.},
    issn = {1931-9401},
    doi = {10.1063/5.0294020},
    url = {https://doi.org/10.1063/5.0294020}
}

@article{PhysRevLett.132.190001,
 _title = {Essay: Quantum Sensing with Atomic, Molecular, and Optical Platforms for Fundamental Physics},
  author = {Ye, Jun and Zoller, Peter},
  journal = {Phys. Rev. Lett.},
  volume = {132},
  issue = {19},
  pages = {190001},
  numpages = {11},
  year = {2024},
  month = {May},
  publisher = {American Physical Society},
  doi = {10.1103/PhysRevLett.132.190001},
  url = {https://link.aps.org/doi/10.1103/PhysRevLett.132.190001}
}

@article{RevModPhys.89.035002,
 _title = {Quantum sensing},
  author = {Degen, C. L. and Reinhard, F. and Cappellaro, P.},
  journal = {Rev. Mod. Phys.},
  volume = {89},
  issue = {3},
  pages = {035002},
  numpages = {39},
  year = {2017},
  month = {Jul},
  publisher = {American Physical Society},
  doi = {10.1103/RevModPhys.89.035002},
  url = {https://link.aps.org/doi/10.1103/RevModPhys.89.035002}
}

@article{science.1146006,
author = {N. Reyren  and S. Thiel  and A. D. Caviglia  and L. Fitting Kourkoutis  and G. Hammerl  and C. Richter  and C. W. Schneider  and T. Kopp  and A.-S. Rüetschi  and D. Jaccard  and M. Gabay  and D. A. Muller  and J.-M. Triscone  and J. Mannhart },
_title = {Superconducting Interfaces Between Insulating Oxides},
journal = {Science},
volume = {317},
number = {5842},
pages = {1196-1199},
year = {2007},
doi = {10.1126/science.1146006},
URL = {https://www.science.org/doi/abs/10.1126/science.1146006},
abstract = {At interfaces between complex oxides, electronic systems with unusual electronic properties can be generated. We report on superconductivity in the electron gas formed at the interface between two insulating dielectric perovskite oxides, LaAlO3 and SrTiO3. The behavior of the electron gas is that of a two-dimensional superconductor, confined to a thin sheet at the interface. The superconducting transition temperature of ≅ 200 millikelvin provides a strict upper limit to the thickness of the superconducting layer of ≅ 10 nanometers.}}

@article{Xu2014,
	abstract = {The recent emergence of two-dimensional layered materials ---in particular the transition metal dichalcogenides ---provides a new laboratory for exploring the internal quantum degrees of freedom of electrons and their potential for new electronics. These degrees of freedom are the real electron spin, the layer pseudospin, and the valley pseudospin. New methods for the quantum control of the spin and these pseudospins arise from the existence of Berry phase-related physical properties and strong spin--orbit coupling. The former leads to the versatile control of the valley pseudospin, whereas the latter gives rise to an interplay between the spin and the pseudospins. Here, we provide a brief review of both theoretical and experimental advances in this field.},
	author = {Xu, Xiaodong and Yao, Wang and Xiao, Di and Heinz, Tony F.},
	date = {2014/05/01},
	date-added = {2026-01-27 15:36:29 +0800},
	date-modified = {2026-01-27 15:36:29 +0800},
	doi = {10.1038/nphys2942},
	id = {Xu2014},
	isbn = {1745-2481},
	journal = {Nature Physics},
	number = {5},
	pages = {343--350},
	_title = {Spin and pseudospins in layered transition metal dichalcogenides},
	url = {https://doi.org/10.1038/nphys2942},
	volume = {10},
	year = {2014},
}

@article{science.aat5975,
author = {Paul R. C. Kent  and Gabriel Kotliar },
_title = {Toward a predictive theory of correlated materials},
journal = {Science},
volume = {361},
number = {6400},
pages = {348-354},
year = {2018},
doi = {10.1126/science.aat5975},
URL = {https://www.science.org/doi/abs/10.1126/science.aat5975},
_eprint = {https://www.science.org/doi/pdf/10.1126/science.aat5975},
abstract = {Correlated electron materials display a rich variety of notable properties ranging from unconventional superconductivity to metal-insulator transitions. These properties are of interest from the point of view of applications but are hard to treat theoretically, as they result from multiple competing energy scales. Although possible in more weakly correlated materials, theoretical design and spectroscopy of strongly correlated electron materials have been a difficult challenge for many years. By treating all the relevant energy scales with sufficient accuracy, complementary advances in Green’s functions and quantum Monte Carlo methods open a path to first-principles computational property predictions in this class of materials.}}

@article{RevModPhys.73.33,
  _title = {Quantum Monte Carlo simulations of solids},
  author = {Foulkes, W. M. C. and Mitas, L. and Needs, R. J. and Rajagopal, G.},
  journal = {Rev. Mod. Phys.},
  volume = {73},
  issue = {1},
  pages = {33--83},
  numpages = {0},
  year = {2001},
  month = {Jan},
  publisher = {American Physical Society},
  doi = {10.1103/RevModPhys.73.33},
  url = {https://link.aps.org/doi/10.1103/RevModPhys.73.33}
}

@article{RevModPhys.68.13,
  _title = {Dynamical mean-field theory of strongly correlated fermion systems and the limit of infinite dimensions},
  author = {Georges, Antoine and Kotliar, Gabriel and Krauth, Werner and Rozenberg, Marcelo J.},
  journal = {Rev. Mod. Phys.},
  volume = {68},
  issue = {1},
  pages = {13--125},
  numpages = {0},
  year = {1996},
  month = {Jan},
  publisher = {American Physical Society},
  doi = {10.1103/RevModPhys.68.13},
  url = {https://link.aps.org/doi/10.1103/RevModPhys.68.13}
}

@article{AFQMC,
author = {Motta, Mario and Zhang, Shiwei},
_title = {Ab initio computations of molecular systems by the auxiliary-field quantum Monte Carlo method},
journal = {WIREs Computational Molecular Science},
volume = {8},
number = {5},
pages = {e1364},
keywords = {ab initio methods, auxiliary-field quantum Monte Carlo, back-propagation, computational quantum chemistry, constrained path approximation, electronic structure, importance sampling, phase problem, phaseless approximation, quantum many-body computation, quantum Monte Carlo methods, sign problem},
doi = {https://doi.org/10.1002/wcms.1364},
url = {https://wires.onlinelibrary.wiley.com/doi/abs/10.1002/wcms.1364},
year = {2018}
}

\end{document}